\documentclass[fleqn,usenatbib]{mnras}

\usepackage{newtxtext,newtxmath}

\usepackage[T1]{fontenc}

\DeclareRobustCommand{\VAN}[3]{#2}
\let\VANthebibliography\thebibliography
\def\thebibliography{\DeclareRobustCommand{\VAN}[3]{##3}\VANthebibliography}

\usepackage[utf8]{inputenc}
\usepackage{natbib}
\usepackage{amsmath}
\usepackage{graphicx}
\usepackage{txfonts}
\usepackage{hyperref}
\usepackage{verbatim}
\usepackage{xspace}
\usepackage{color}
\usepackage{breqn}
\usepackage{multicol,lipsum}
\usepackage{etoolbox}
\usepackage{enumitem}
\usepackage{xcolor}
\usepackage{caption}
\usepackage{subcaption}

\newcommand\rg{$r_g~$}

\newcommand\sgra{Sgr A$^\ast$}

\title[Black hole accretion characterization with invariants]{Characterization of black hole accretion through image moment invariants}

\author[Jim\'{e}nez-Rosales, Yfantis, Mo\'scibrodzka, et al.]{
A. Jim\'{e}nez-Rosales$^{1}$\thanks{a.jimenez@science.ru.nl}, A. I. Yfantis$^{1}$, M. Mo\'scibrodzka$^{1}$,
J. Dexter$^{2}$.
\\
$^{1}$Department of Astrophysics, IMAPP, Radboud University, 6500 GL Nijmegen, The Netherlands\\
$^{2}$JILA and Department of Astrophysical and Planetary Sciences, University of Colorado, Boulder, CO 80309, USA\\
}

\date{Accepted XXX. Received YYY; in original form ZZZ}

\pubyear{2023}

\begin{document}
\label{firstpage}
\pagerange{\pageref{firstpage}--\pageref{lastpage}}
\maketitle

\begin{abstract}
We apply image moment invariant analysis to total intensity and polarimetric images calculated from general relativistic magnetohydrodynamic simulations of accreting black holes. We characterize  different properties of the models in our library by their invariant distributions and their evolution in time. We show that they are highly sensitive to different physical effects present in the system which allow for model discrimination.
We propose a new model scoring method based on image moment invariants that is uniformly applicable to total intensity and polarimetric images simultaneously. 
The method does not depend on the type of images considered and its application to other non-ring like images (e.g., jets) is straight forward.

\end{abstract}




\section{Introduction}
\label{sec:intro}

Accretion onto astrophysical compact sources is the engine that drives some of the most powerful and luminous events in the universe. 
In the case of supermassive black holes (SMBH), the study of magnetized accretion flows at near-horizon regions allows for the inference of the emission region nature and immediate environment around the source, as well as the properties of the central object.

Due to their relatively large size on-sky as viewed from Earth, two important 
SMBH are Sagittarius A$^\ast$ (\sgra) at the center of our Galaxy and Messier 87$^\ast$ (M87$^\ast$), at the center of the giant elliptical galaxy M87. Both are accreting matter at strongly sub-Eddington rates. Many analytic and semi-analytic models have been used to model these low-luminosity accretion systems \citep[e.g.][]{ichimaru1977,rees1982,narayan1995,reynolds1996,blandford1999,falcke2000jet,falcke2000shadow,melia1998,bromley2001,yuan2003,broderick2006,yuan2014}. 

In a fully numerical context, the flows onto SMBHs can be commonly studied using general relativistic magneto hydrodynamic (GRMHD) simulations \citep[e.g.][]{devilliers2003, gammie2003, tchekhovskoy2011,dibi2012,sadowski2015,porth2017,tchekhovskoy2019}, where the accretion process is initiated and self-consistently evolved as a result of turbulence and plasma instabilities, e.g. the magneto-rotational instability \citep{balbus1991}, at scales within a few gravitational radii from the compact object. 

In a conservative framework, the general relativistic equations of magnetohydrodynamics (GRMHD) are solved in a particular spacetime and evolved according to conservation laws for mass, energy-momentum and the Maxwell equations. 
A notable result from these calculations is that the magnetic field evolution and accretion can naturally produce powerful
relativistic outflows by tapping into the black hole spin energy in the form of jets \citep{blandford1977} or winds \citep{blandfordpayne}.

In these hot under-luminous accretion flows, the dynamics of the disk are set by the heavy ions, while the near-horizon emission is dominated by synchrotron radiation emitted by the lighter relativistic electrons. 
The latter are typically not directly simulated in single-fluid GRMHD simulations and, by employing physical arguments (e.g. charge neutrality), the electron population characteristics can be inferred. Only the electron internal energy and electron distribution function remain unconstrained.
One approach is to assume a thermal distribution function and assign an electron temperature in post-process. There are many choices of this, from a constant fraction of the internal energy of the ions \citep[][]{goldston2005}, or as a function of the magnetic field strength \citep[e.g.][]{moscibrodzka2013,moscibrodzka2014,chan2015image}, to other kinetic prescriptions that include anisotropic viscosity \citep[][]{sharma2007}. Alternatively, different prescriptions can be included in the GRMHD simulation to infer the properties of the electrons \citep[e.g.][]{vaidya2018}. In particular, new algorithms have been developed which allow for a self-consistent evolution of the electron thermal energy with that of the MHD fluid  \citep[e.g.][]{ressler2015,ressler2017,gold2017,chael2018,chael2019}. These are based on electron heating prescriptions resulting from particle in-cell simulations \citep[e.g. turbulent cascade scenarios or magnetic reconnection events;][]{howes2010,rowan2017,werner2018,kawazura2019,zhdankin2019}.

A variety of observations of Sgr~A* and M87* (including total intensity, image morphology, polarization and others), can be successfully reproduced with these models \citep[e.g.][]{noble2007,dexter2009,moscibrodzka2009,dexter2010,shcherbakov2012,moscibrodzka2014,chan2015var,gold2017,dexter2020,jimenez-rosales2020,EHTPaperI,EHTPaperVIII,EHTSgraPaperI}. However, much is still unknown about the nature of the systems and many degeneracies are still present in the models. This leaves room for much improvement and motivates the development of tools and techniques that exploit the richness of the observations to constrain the models better \citep[e.g.][]{johnson2017,johnson2018,johnson2020,jimenez-rosales2018,jimenez-rosales2021,palumbo2020,narayan2021, ricarte2021,ricarte2023}. 

The first successful observations of \sgra and M87$^\ast$ of the Event Horizon Telescope (hereafter EHT) made with Very Long Baseline Interferometry (VLBI) at an observing wavelength of 1.3 mm \citep[230 GHz;][]{EHTPaperI,EHTPaperII,EHTPaperIII,EHTPaperIV,EHTPaperV,EHTPaperVI,EHTPaperVII,EHTPaperVIII,EHTSgraPaperI,EHTSgraPaperII,EHTSgraPaperIII,EHTSgraPaperIV,EHTSgraPaperV,EHTSgraPaperVI}, have opened a new window to studying these sources using the properties of the image alone and motivates the approach of this work to tackle the problem from an image processing and computer vision perspective.

Some works have already explored computer vision techniques, namely neural networks, and their application to parameter estimation analysis of EHT data and models \citep[e.g.][]{gucht2020,yao-yu2020,yao-yu2021,qiu2023}. Here we take a different approach and focus on image moment invariants (IMI).

IMI are quantities constructed from image moments that exhibit the property of not changing under a particular affine transformation, such as translation, rotation or scaling of the image; which allows for a convenient and powerful way to characterize images. 

IMI were introduced by \citet{hu1962}, as a new tool for pattern recognition. Since then, many advances have taken place in the form of improvements and generalizations of Hu’s seven famous invariants. Their application has been very fruitful in many other areas such as in medical imaging in two and three dimensions \citep[e.g.][]{mangin2004, ng2006, xu2008, li2017} and feature extraction \citep[e.g.][]{flusser1993,yang2011,yang2015,yang2017,yang2018,zhang2015}.  

In this work we explore how geometric IMI can be used for model comparison and parameter estimation of accretion onto SMBH, using high resolution radio and millimeter wavelength images of astrophysical objects made from data collected by VLBI observations such as, e.g., EHT. When comparing models to the VLBI images, the advantages of using invariants (e.g. under translation or rotation) becomes obvious, not only because the position angle of the images is an unconstrained model parameter, but also because VLBI image reconstructions do not have an absolute reference frame, such that reconstructed images may be shifted off centre. 

VLBI data are usually recorded in full polarization which permits the reconstruction of images in all four Stokes parameters (${\mathcal I,Q,U,V}$) at multiple wavelengths. 
The goal of this work is to develop a model comparison method that can be applied uniformly to all type of images. 

This paper is structured as follows. Section \ref{sec:numerical_setup} gives the information on the numerical simulations used in this work as well as the ray-tracing techniques used to generate a library of images of gas around a black hole. In Section~\ref{sec:moment_invariants} we present the mathematical background for the image moments and invariants with respect to translation, rotation and scaling. Sections~\ref{sec:results1} and ~\ref{sec:results2} present the main analysis of image invariants and scoring ideas. In Section~\ref{sec:time}, we study the time-dependent behavior of these quantities. Lastly, Section~\ref{sec:discussion} contains discussion and conclusions of our work.

\section{Black hole image library}~\label{sec:numerical_setup}

\begin{table}
	\centering
	\caption{3-D GRMHD simulations used in this work to create library of black hole images.}
	\label{table:models}
	\begin{tabular}{lccc} 
		\hline
		Model  & $e^-$ heating & $i$ & $a$   \\
		\hline
		MAD  & TB, RC & $25\deg$ & 0.0, 0.5, 0.9375  \\
		SANE & TB, RC & $25\deg$ & 0.0, 0.5, 0.9375 \\
		\hline
	\end{tabular}
\end{table}

The black hole image library we use is calculated as described in the following. 

We use a subset of 3-D GRMHD, long-duration simulations of black hole accretion described in \citet{dexter2020}. 
The simulations have been carried out with the HARMPI\footnote{https://github.com/atchekho/harmpi} code \citep{tchekhovskoy2019}, using conservative MHD in a fixed Kerr space–time.

The initial condition consists of a Fishbone–Moncrief torus
\citep{fishbone1976} with inner radius at $r_{\rm in} = 12~M$ (in geometric units) 
and pressure maximum radius $r_{\rm max} = 25~M$. We consider three values of dimensionless black hole spin $a = 0,0.5$ and $0.9375$. The torus is threaded with a single poloidal loop of magnetic field whose radial profile can provide either a highly saturated (magnetically arrested disc, MAD) or relatively modest (standard and normal evolution, SANE) magnetic flux.
For more details see \citet{dexter2020}.

The GRMHD algorithm includes a self-consistent heating prescription of the electron internal energy density in pair with that of the single MHD fluid as implemented by \citet{ressler2015}. In this case, the fluid receives a fraction of the local dissipated energy according to a chosen sub-grid prescription motivated by kinetic calculations \citep[e.g.][]{howes2010, rowan2017, werner2018, kawazura2019}. Here, we explore two electron heating prescriptions out of the set of \citet{dexter2020}: turbulent heating (TB) based on gyrokinetic theory \citep{howes2010} and magnetic reconnection (RC) from particle-in cell simulations \citep{werner2018}. We also assume a composition of pure ionized hydrogen \citep{wong2022}.

To predict observational appearance (images) of the GRMHD models we carry out radiative transfer simulations. We post-process GRMHD simulations using the general relativistic ray-tracing public code GRTRANS\footnote{https://github.com/jadexter/grtrans} \citep{dexter2009,dexter2016}. We assume a fast-light approximation.

With the emission mechanism set to synchrotron emission, we compute the full Stokes vector ($I,Q,U,V$), which characterizes the properties of the polarized light. 
The electron temperature $T_e$ is obtained from the GRMHD electron internal energy density $u_e$ according to $k_B T_{\rm e} = (\gamma_{\rm e}-1)m_{\rm p} u_{\rm e}/\rho$, where $k_B$ is the Boltzmann constant, $\gamma_{\rm e} = 4/3$ the adiabatic index for relativistic electrons, $m_{\rm p}$ the proton mass and $\rho$ the density.  
The mass of the black hole sets the time scale of the system. We use a value of $4 \times 10^6 M_{\odot}$, where $M_{\odot}$ is the solar mass unit, consistent with that of Sgr ~A*. The mass accretion rate is scaled to match the observed flux density at 230~GHz of \sgra \citep[2.5 Jy; e.g.][]{dexter2014,bower2015,EHTSgraPaperI}. 
We calculate images at 230~GHz at an inclination of $i=25\deg$, where the viewing angle is motivated by recent observations of Sgr~A* \citep{gravity2020,wielgus2022,EHTSgraPaperV}. 
The images resolution is always $192\times192$ pixels over a 42~\rg ($\sim210~\mu \rm{as}$) field-of-view. We blur the images with a $20~\mu as$ Gaussian, consistent with the characteristic imaging resolution of the EHT. 

\begin{figure}
\centering
\includegraphics[trim={0cm 0cm 0cm 0cm},clip=True,width=0.49\columnwidth]{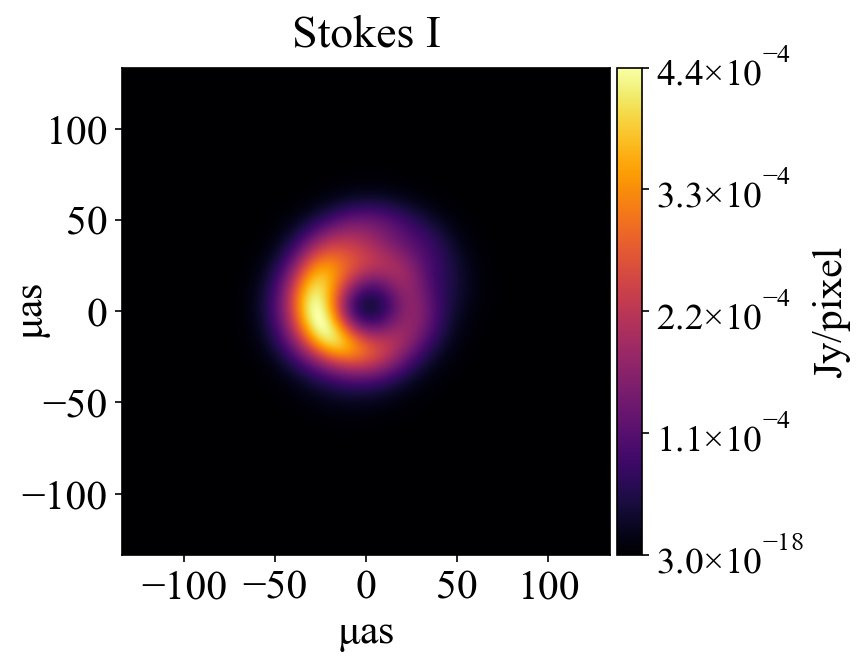}
\includegraphics[trim={0cm 0cm 0cm 0cm},clip=True,width=0.49\columnwidth]{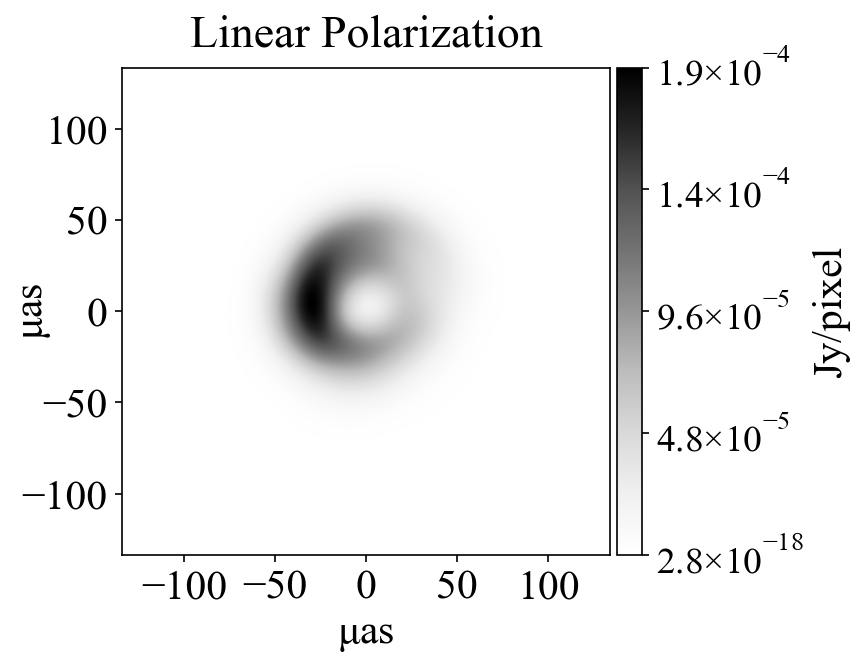}
\includegraphics[trim={0cm 0cm 0cm 0cm},clip=True,width=0.49\columnwidth]{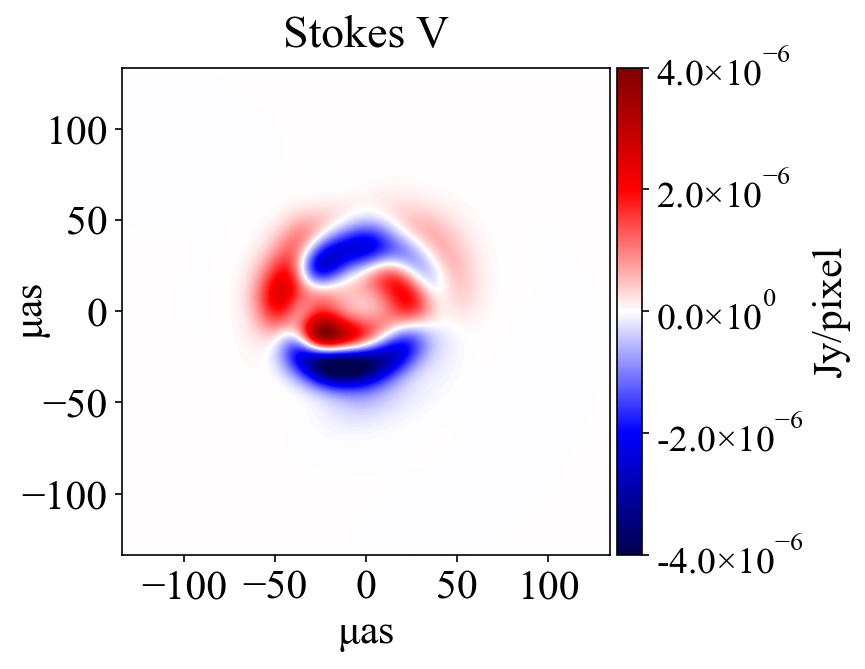}
\vspace{0.0cm}
\caption{Example polarized images from a model snapshot in our library. The images have been blurred with a $20\mu$as Gaussian, characteristic of the EHT resolution. Top left: Total intensity (Stokes I). Top right: Linear polarization. Bottom: Stokes V. }
\label{fig:example_frames}
\end{figure}

Example of blurred polarized images from one frame from our library are shown in Fig.~\ref{fig:example_frames}. 
It can be seen that the Stokes $I$ and linear polarization ($LP\equiv\sqrt{Q^2+U^2}$) images are typically dominated by ring-like features. This is true for this type of low-luminosity systems due to their low optical thickness and large geometric thickness. The Stokes $V$ images show more structure and can have both positive and negative values. A summary of the parameters used for the models is shown in Table~\ref{table:models}.
Our total image sample consists of $150$ frames per model, spanning a range of $1500~M$.

\section{Image Moment Invariants}~\label{sec:moment_invariants}

In 2-D, the moment of order $pq$ of a function $f(x,y)$ is defined as:
\begin{equation}
\label{eq:moment_def}
m_{pq} = \int_{\infty}\int_{\infty}\Phi_{pq}(x,y)f(x,y)~dx~dy;
\end{equation}
\noindent where $\Phi_{pq}$ corresponds to a particular set of basis functions. The indices $p,~q$ usually denote, respectively, the degree of order of the coordinates $x$ and $y$ as stated within $\Phi$. In the case of a 2-D image, the function $f(x,y)$ represents the value of a pixel $(x,y)$ and the integrals become discrete sums over the image extent.

In general, moments can be constructed using variety of basis functions depending on the application \citep[e.g. geometric, Gauss-Hermite; see][]{flusser1993,yang2018}.
In this work we use a geometric basis, where the set of functions consists of polynomials of order $n\geq0$, where the latter is an integer.

Some image moments have a well understood physical interpretation. For example, $m_{00}$ is associated to the total flux of the image, while $(m_{10}/m_{00},~m_{01}/m_{00})$ represents the centroid of the image. As the order of the moment increases, however, assigning a physical meaning becomes a difficult task.

Invariant quantities under affine transformations can be constructed using image moments. 

In a geometric basis, the \textit{centralized moments} are translation invariant and are defined as:
\begin{equation}
    \mu_{pq} = \sum_x\sum_y (x-x_0)^p(y-y_0)^q f(x,y) dx dy;
\end{equation}
\noindent where $(x_0,y_0)$ is the centroid of the image and is calculated following Eq.~\ref{eq:moment_def} with $\Phi_{pq}(x,y) = x^p~y^q$. 

The set of $\mu_{pq}$ can be made scale invariant (i.e. with respect to image size) by
\begin{equation}
\label{eq:scale_mom}
    \eta_{pq} = \frac{ \mu_{pq} }{ \text{abs}\left( \mu_{00}^{\left[1+\frac{p+q}{2}\right]} \right) }.
\end{equation}
\noindent In this work, the intensity function $f(x,y)$ can represent any of the polarized quantities $I,~LP,~\&~V$. We note that by taking the absolute value in the denominator, we have extended the usual definition of Eq.~(\ref{eq:scale_mom}) to allow negative valued pixels, as can be the case for Stokes V (see Appendix~\ref{appdx:verification}).

In addition to these transformations, \citet{hu1962} and \citet{flusser1993} showed that the following moment combinations are rotationally invariant \citep{hu1962,flusser2000}:

\begin{equation}\label{eq:hu_invariants}
\begin{aligned}
HF_0 &= \eta_{20} + \eta_{02}\\
HF_1 &=  (\eta_{20}-\eta_{02})^2+4\eta^2_{11}\\
 HF_2 &= (\eta_{30}-3\eta_{12})^2 + (3\eta_{21}-\eta_{03})^2 \\
    HF_3 &= (\eta_{30}+\eta_{12})^2+(\eta_{21}+\eta_{03})^2 \\
    HF_4 &= (\eta_{30}-3\eta_{12})(\eta_{30}+\eta_{12})[(\eta_{30}+\eta_{12})^2-3(\eta_{21}+\eta_{03})^2]+ \\
    & (3\eta_{21}-\eta_{03})(\eta_{21}+\eta_{03})[3(\eta_{30}+\eta_{12})^2-(\eta_{21}+\eta_{03})^2]\\
    HF_5 &= (\eta_{20}-\eta_{02})[(\eta_{30}+\eta_{12})^2-(\eta_{21}+\eta_{03})^2]+\\
    & 4\eta_{11}(\eta_{30}+\eta_{12})(\eta_{21}+\eta_{03}) \\
    HF_6 &= (3\eta_{21}-\eta_{03})(\eta_{30}+\eta_{12})[(\eta_{30}+\eta_{12})^2-3(\eta_{21}+\eta_{03})^2]-\\
    & (\eta_{30}-3\eta_{12})(\eta_{21}+\eta_{03})[3(\eta_{30}+\eta_{12})^2-(\eta_{21}+\eta_{03})^2] \\
    HF_7 &= \eta_{11}[(\eta_{30}+\eta_{12})^2-(\eta_{03}+\eta_{21})^2]-\\
    & (\eta_{20}-\eta_{02})(\eta_{30}+\eta_{12})(\eta_{03}+\eta_{21}) 
\end{aligned} 
\end{equation}

In the rest of this work we will refer to this set as the ``HF'' invariants. Here, $HF_k;~k\leq 6$, are the "original" invariants proposed by \citet{hu1962}. $HF_7$ was later added by \citet{flusser2000}. Though Eqs.~\ref{eq:hu_invariants} are neither complete nor independent ($HF_2 = (HF_4^2+HF_6^2)/HF_3^3$), they have been used to successfully extract and characterize image features and are widely used in image recognition algorithms (see Section~\ref{sec:intro}).

An analogy could be made between $\text{abs}(HF_0)$, and the moment of inertia of an object about an axis. In this case, the pixels' intensities would be analogous to the object's density and the rotation axis to the image's centroid.  
An interesting property of the $HF$ invariants is that while the $HF_k;~k\leq5$, are reflection symmetric, $HF_6$ and $HF_7$ are skew invariant, which could allow for a distinction of mirrored images in a collection of otherwise identical images. For more details on how
the invariants change under certain image transformations, see Appendix~\ref{appdx:properties}.

It is clear that the complicated functional form of the invariants prevents a clear interpretation of what each quantity is measuring. We investigated this further but were unsuccessful in identifying a clear tendency of how the HF change with respect to different images (see Appendix.~\ref{appdx:min_max_diff}).

In what follows we will explore their general behavior when characterising our black hole library and their dependence on physical effects.

\section{Invariant distributions as a function of model physical parameters}\label{sec:results1}

We first explore the sensitivity of the image moment invariants to the physical model parameters. We present and discuss the distributions of the invariants computed per polarized image quantity $(I,~LP,~V)$ for different magnetizations, black hole spins and electron-heating mechanisms, separately.

\subsection{Invariants based on total intensity, Stokes \textit{I}}

\begin{figure*}
\centering
\begin{subfigure}{1.0\textwidth}
    \centering
    {\large Stokes~$I$} \\ \vspace{0.2cm}
    \includegraphics[trim={0cm 0cm 0cm 5.4cm},clip=True,width=1\columnwidth]{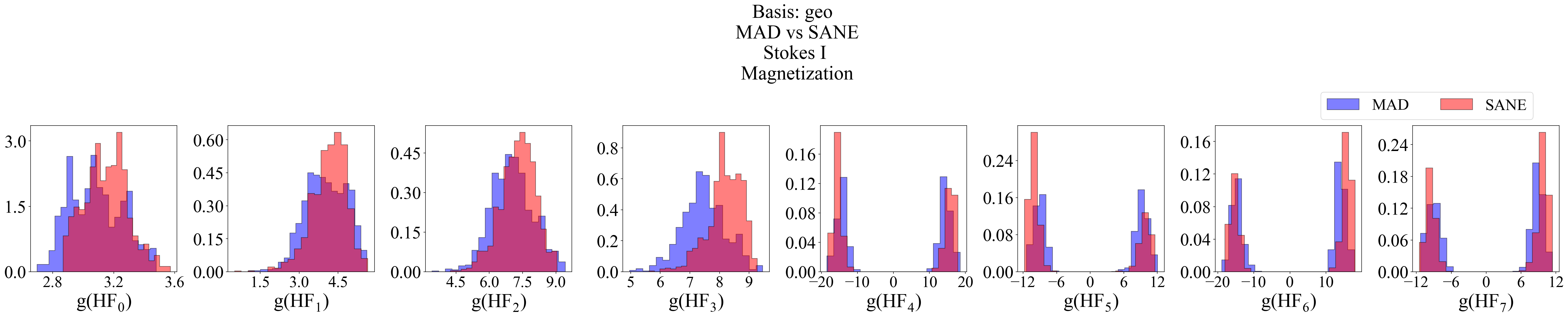}
    \includegraphics[trim={0cm 0cm 0cm 5.4cm},clip=True,width=1\columnwidth]{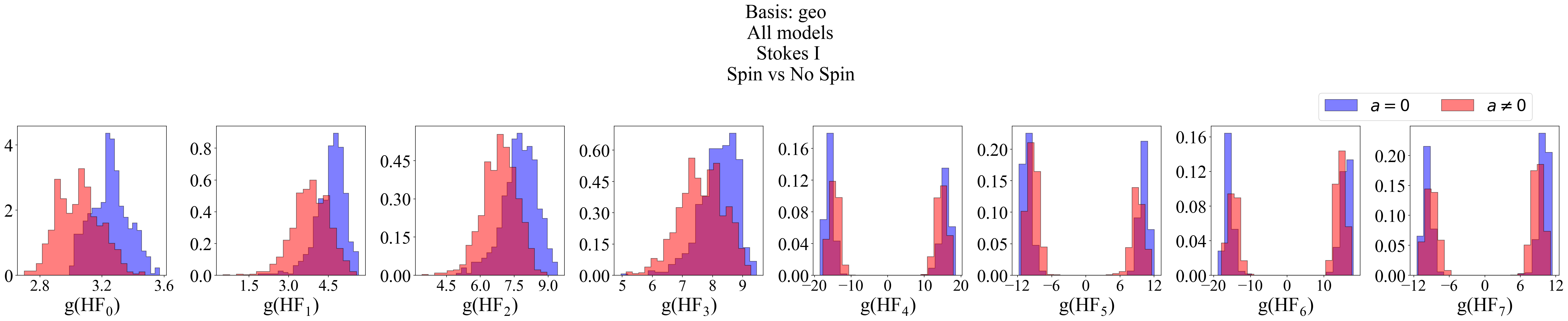}
    \includegraphics[trim={0cm 0cm 0cm 5.4cm},clip=True,width=1\columnwidth]{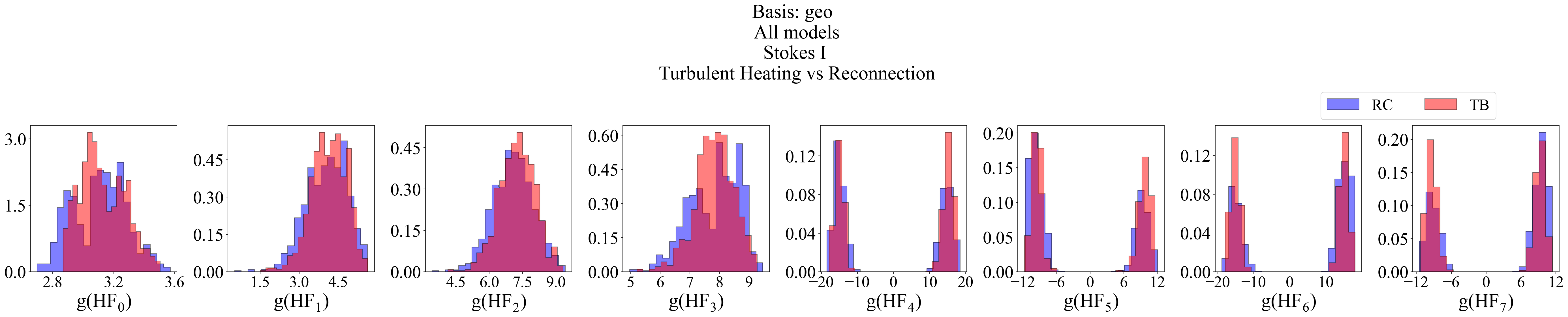}
\end{subfigure}
\caption{Normalized distributions of $g(HF_k)=\textrm{sign}(HF_k)\log|HF_k|$ of Stokes $I$ images as a function of various physical effects. Columns indicate a different HF invariant. Rows indicate in order: magnetization (top), spin (middle), electron heating (RC vs TB, bottom). Each row marginalizes over all non-divided parameters. }
\label{fig:distr_allI}
\end{figure*}

Fig.~\ref{fig:distr_allI} shows the distributions of $g(HF_k)=\textrm{sign}(HF_k)\log(~|HF_k|~)$ for Stokes~$I$, as a function of different physical effects. The first row shows the distributions for total flux, $I$, as a function of magnetization. 
It is interesting to note that for $0\leq k\leq 3$, the $HF_k$ distributions for either the MAD (blue) or the SANE (red) cases, appear to concentrate in one lobe, while cases $HF_k;~k>3$, split into two. Simultaneous sign changes of certain $HF_k$ should be indicative of a degree of mirroring in the images or sign flip (or both, mirroring and sign flip) in case of Stokes V maps (see Appendix~\ref{appdx:properties}).

In the case of the one-lobe distributions, both populations span over a similar order-of-magnitude range, with the MAD extending slightly more to lower values. The MAD distribution shows a relatively higher level of symmetry compared to the SANE, which tend to skew toward low orders of magnitude. The case $HF_0$ shows an interesting behavior of the MAD population, with a bi- or even tri-modal population. The maxima at lower values represents mostly non-zero spin reconnection MAD models, while the one at higher values the models with zero spin and turbulent heating mechanism (Fig.~\ref{fig:distr_huI_sign}). The peak in the middle is a combination of the rest. 
The cases where $HF_k;~k>3$, the bi-modal distributions of the models are similar across the value of $k$. For either magnetization case, each lobe is located at approximately the same distance from the origin as their $\pm$ counterpart. If only $\log(~|HF_k|~)$ was plotted instead, without sign information, the populations would form one continuous lobe with similar shape to the cases where $HF_k;~k\leq 3$. The ``negative'' models would comprise the higher end of their respective $\log(~|HF_k|~)$ distributions. Cross-referencing these panels to the ones in Fig.~\ref{fig:distr_huI_sign}, it can be seen that either family of lobes correspond to different frames of the same model. 

In the case spin vs no spin, shown in the middle row of Fig.~\ref{fig:distr_allI}, the different spin populations appear to be in a one-lobe distribution for $k\leq3$, while the distributions split into two for $k>3$. We found that the $a=0.5$ models lie generally between the $a=0$ and $a=0.94$ and so, in the interest of simplicity and to make the differences between the cases more evident, we decided to separate the models into two categories: zero and non-zero spin, where the latter includes both $a=0.5$ and $a=0.94$. The $a=0$ population (blue) is skewed to low orders of magnitude, as opposed to the $a\neq 0$ population (red) which looks relatively symmetric. Introducing spin to a non-spinning population appears introduce a translation of the distribution to lower orders of magnitude, without preserving the shape. In the case of $HF_0$, the lower end of the spinning populations is dominated by the MAD, while the $a=0$ SANE dominates the higher end of the non-spinning distribution (Fig.~\ref{fig:distr_huI_sign}).

The bottom row of Fig.~\ref{fig:distr_allI}, shows the effect of changing the electron heating mechanism of the invariant population. Turbulent heating (TB, red) displays what is mostly a one-lobe distribution with a slight skewness to low values. Reconnection (RC, blue) concentrates as well in one lobe and spans about the same range as the TB population. Occasionally, it shows a two lobe distribution (e.g. $HF_0,~HF_3$). In these cases, the lobe located at lower values is consistent with non-zero spin reconnection models, the majority of them being MAD (see bottom panel of Fig.~\ref{fig:distr_huI_sign}).

\subsection{Invariants based on linear polarization, \textit{LP}}

\begin{figure*}
\centering
\begin{subfigure}{1.0\textwidth}
    \centering
    {\large Linear Polarization} \\ \vspace{0.2cm}
    \includegraphics[trim={0cm 0cm 0cm 5.4cm},clip=True,width=1\columnwidth]{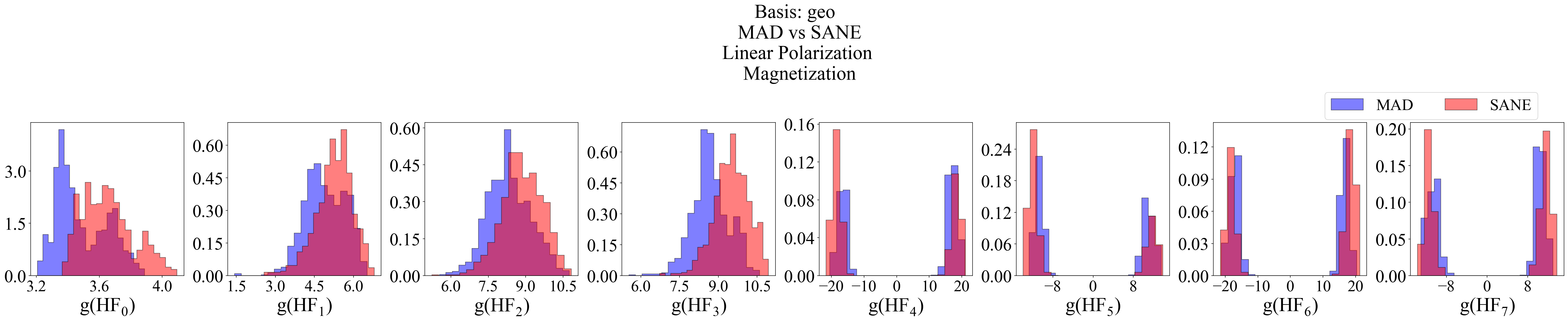}
    \includegraphics[trim={0cm 0cm 0cm 5.4cm},clip=True,width=1\columnwidth]{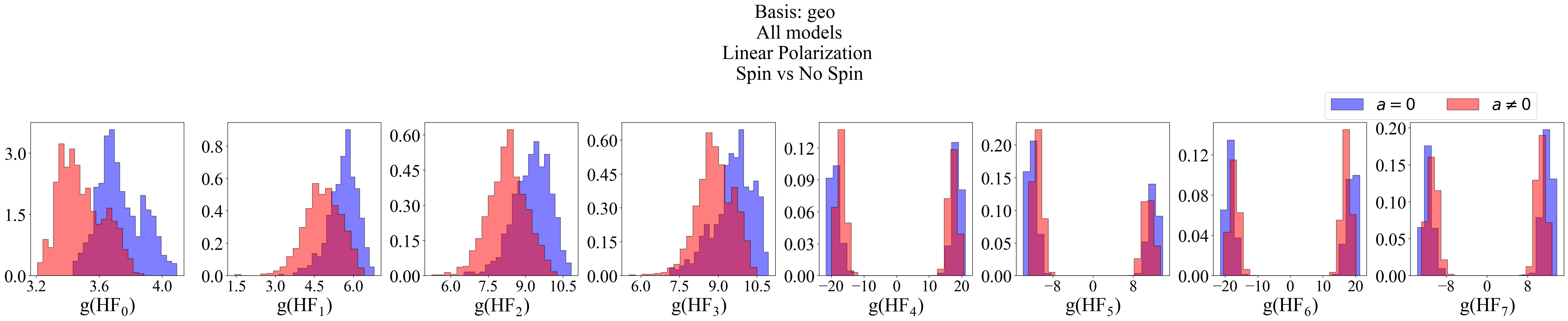}
    \includegraphics[trim={0cm 0cm 0cm 5.4cm},clip=True,width=1\columnwidth]{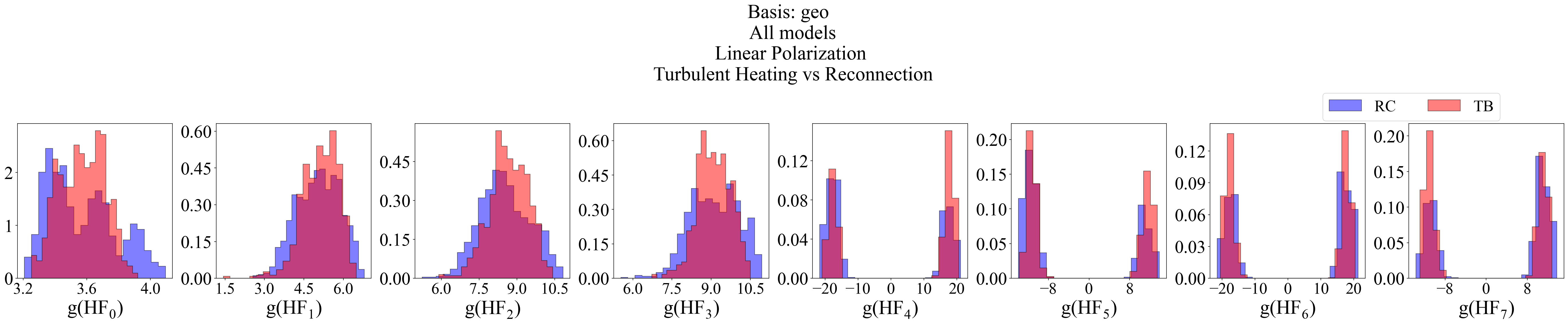}
\end{subfigure}
\caption{ Similar as Fig.~\ref{fig:distr_allI} but for $LP$.  }
\label{fig:distr_allLP}
\end{figure*}

Similarly to Fig.~\ref{fig:distr_allI}, the distributions in Fig.~\ref{fig:distr_allLP} show two distinct sets of shapes for the cases $k\leq 3$ and $k>3$ regardless of the physical effect dependency. 

The distributions for all MAD and SANE models are shown in the top row of Fig.~\ref{fig:distr_allLP}. The distinction between MAD (blue) and SANE (red) is more evident. For $0<k\leq 3$, both distributions are concentrated in one lobe. The MAD population, however, is relatively symmetric and is concentrated toward lower order of magnitude values. On the other hand, the SANE distributions concentrate towards the higher range of the span and present a tail to lower values. There is considerable overlap between both distributions. 
For $k=0$, both distributions appear to be bimodal. Complementing this information with that from the middle and bottom panels of Fig.~\ref{fig:distr_huLP_sign}, it can be seen that the smaller lobe of the SANE distribution is made up by zero spin, reconnection models. 
For $k > 3$, either distribution splits into two lobes, each comprised of different frames of the same models and each located at approximately the same distance from the origin as their $\pm$ counterpart. If $\log(~|HF_k|~)$ was plotted, the populations would show similar one-lobe shapes as for the $HF_k;~k\leq3$ cases. 
 
Analyzing the $LP$ distributions from a spin vs no spin perspective (middle row of Fig.~\ref{fig:distr_allLP}), a similar distribution shape to MAD vs SANE is observed across $k$ values. Comparing the relative positions of both distributions, it is clear that a change in spin introduces a translation of the distribution, where an increase in $a$ moves the invariant populations toward lower values. 
    
The bottom row of Fig.~\ref{fig:distr_allLP} shows the effects of a change in electron heating. Regardless of the electron heating, the populations span over a similar domain. 
The SANE RC models seem to make up most of the higher end values while the MAD RC dominate the lower end.

\subsection{Moment invariants based on circular polarization, Stokes \textit{V}}

\begin{figure*}
\centering
\begin{subfigure}{1.0\textwidth}
    \centering
    {\large Stokes~$V$} \\ \vspace{0.2cm}
    \includegraphics[trim={0cm 0cm 0cm 5.4cm},clip=True,width=1\columnwidth]{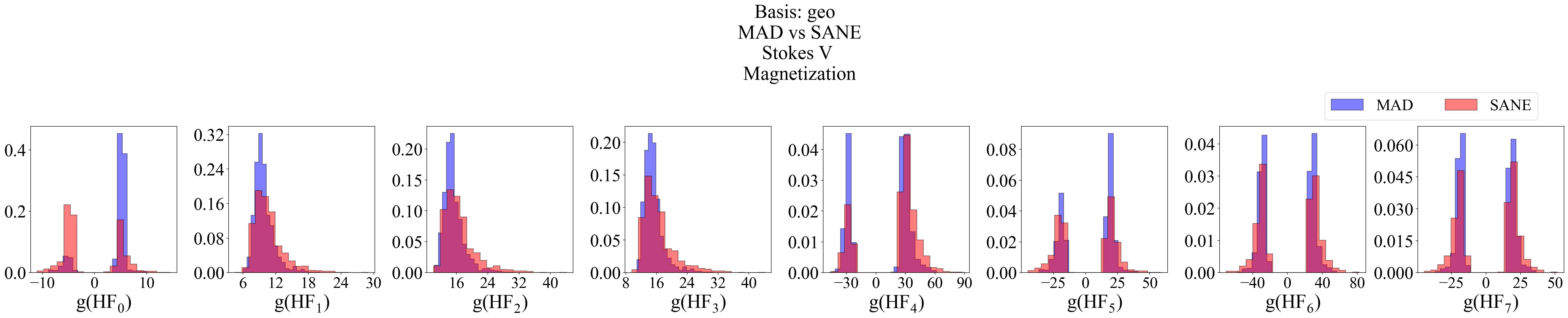}
    \includegraphics[trim={0cm 0cm 0cm 5.4cm},clip=True,width=1\columnwidth]{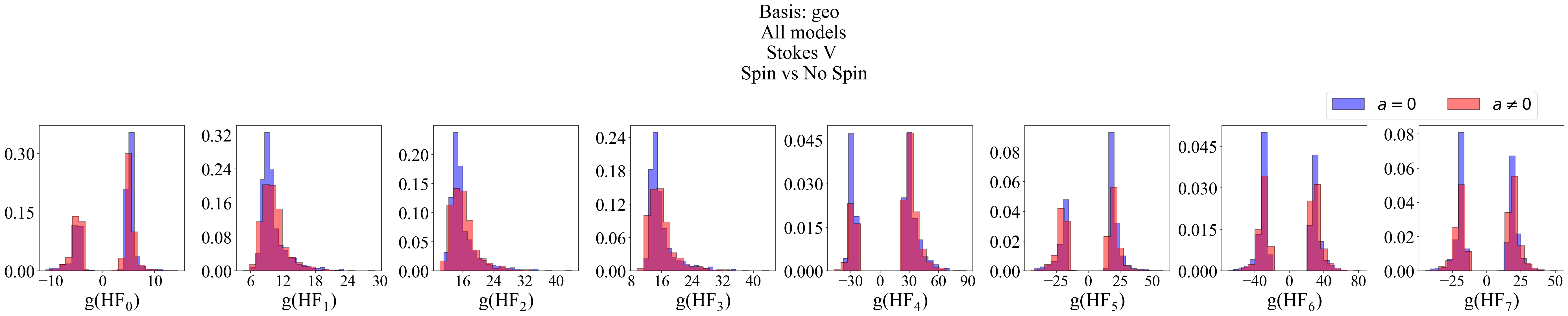}
    \includegraphics[trim={0cm 0cm 0cm 5.4cm},clip=True,width=1\columnwidth]{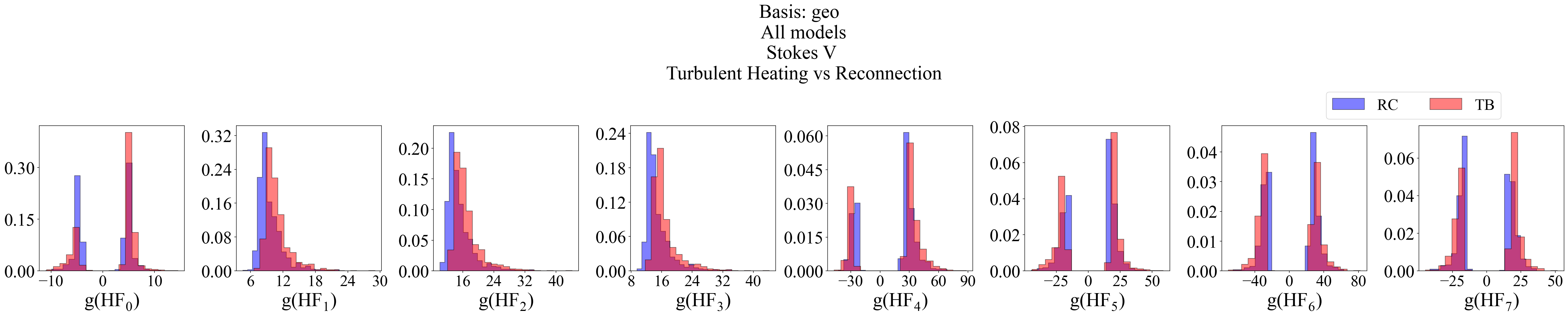}
\end{subfigure}
\caption{ Similar as Fig.~\ref{fig:distr_allI} but for $V$.  }
\label{fig:distr_allV}
\end{figure*}

Similarly to Figs.~\ref{fig:distr_allI} and \ref{fig:distr_allLP}, Fig.~\ref{fig:distr_allV} shows the normalized IMI distributions for circular polarization images. 
 
The shape of the populations is very similar across all physical effects. In the cases $0<k\leq 3$, the overall shape displays a concentration towards the lower end of the span with a tail extending to higher order values. Interestingly, for Stokes $V$ the case $k=0$, shows a two-lobe split distribution of $g(HF_k)$ like the one seen for $k>3$, contrasting with Stokes $I$ and $LP$, where only the latter invariants showed these features. Just as for $I$ and $LP$, the lobes are comprised of different frames of the same models and are located at approximately the same distance from the origin as their $\pm$ counterpart. Once again, if $\log(~|HF_k|~)$ was plotted instead, the populations would concentrate in one lobe, skewed to high values. 

Stokes $V$ does not seem to have a significant sensitivity to physical effects. The only distinguishable difference between any pair of distributions appears to be the location of their maxima. In the case of magnetization, the MAD distribution are more concentrated to one value lower than the SANE. In the case of spin, the $a=0$ distribution has a maxima at relatively lower values than the $a\neq 0$. In both the $a=0$ and $a\neq 0$ cases, the bulk of the distribution appears to be dominated by MAD models.

It appears that Stokes $V$ is most sensitive to a change in electron heating. A change in electron heating results in a slight offset between their maxima, with the RC dominating the lower range of values and being made up almost in its entirety by MAD models (see Fig.~\ref{fig:distr_huV_sign}). The TB population peaks at larger values than the RC and appears to be also dominated by the MAD.

\subsection{Linear vs Circular Polarization Invariants}

\begin{figure*}
\centering
\begin{subfigure}{1.0\textwidth}
    \centering
    {\large Linear Polarization \& Stokes V} \\ \vspace{0.2cm}
    \includegraphics[trim={0cm 0cm 0cm 5.4cm},clip=True,width=1.0\columnwidth]{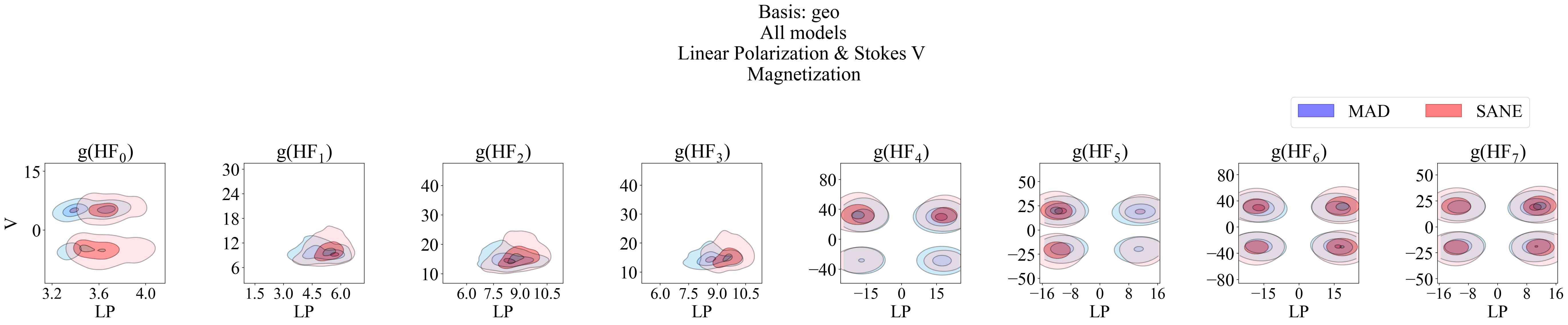}
    \includegraphics[trim={0cm 0cm 0cm 5.4cm},clip=True,width=1.0\columnwidth]{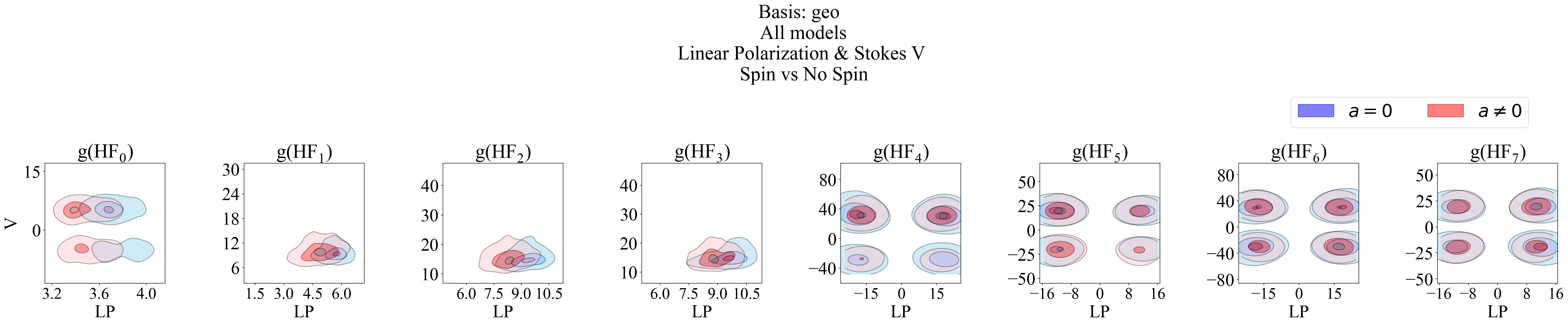}
    \includegraphics[trim={0cm 0cm 0cm 5.4cm},clip=True,width=1.0\columnwidth]{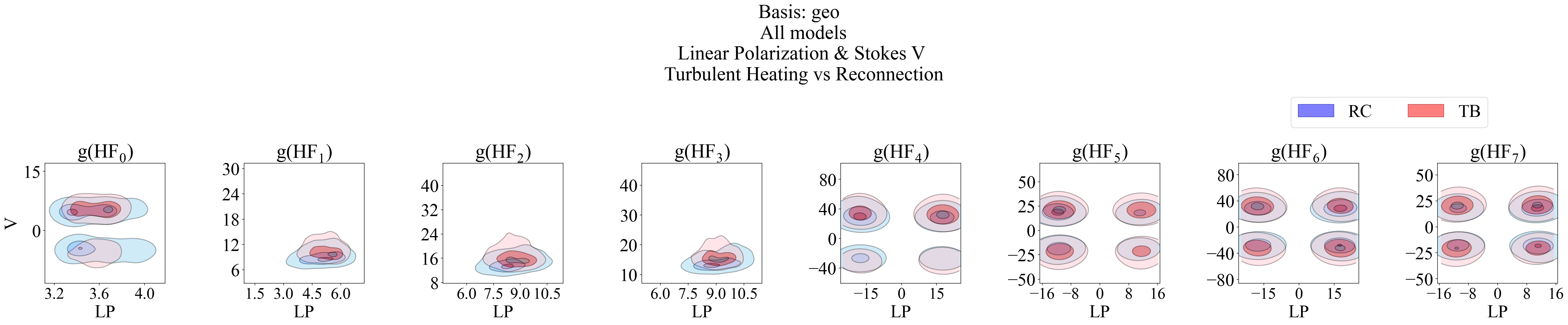}
\end{subfigure}
\caption{ Similar to Fig.~\ref{fig:distr_allI} but for $LP$ \& $V$ together.}
\label{fig:distr_hulp_v_sign}
\end{figure*}

It is also instructive to plot the HF invariants in LP vs V domain. This is shown in Fig.~\ref{fig:distr_hulp_v_sign}

The behavior of the HF invariants remains very similar regardless of the physical effect (magnetization, spin, electron heating). In the case of HF$_0$, both sets of distributions form two separate ``islands'', this is because Stokes V splits into two as is seen in the first column of Fig.~\ref{fig:distr_allV}. 
For HF$_k;~1\leq k \leq3$, the distributions form one island with extended and significant overlap.

In the case of HF$_k;~4\leq k$, since both the $LP$ and Stokes $V$ distributions slips into two, when shown as $V~vs~LP$, this results in four islands where the respective populations overlap almost entirely.

\section{Model scoring procedures using moment invariants}\label{sec:results2}

\begin{figure*}
\centering
\includegraphics[trim={0cm 0cm 0cm 0.5cm},clip=True,width=1.8\columnwidth]{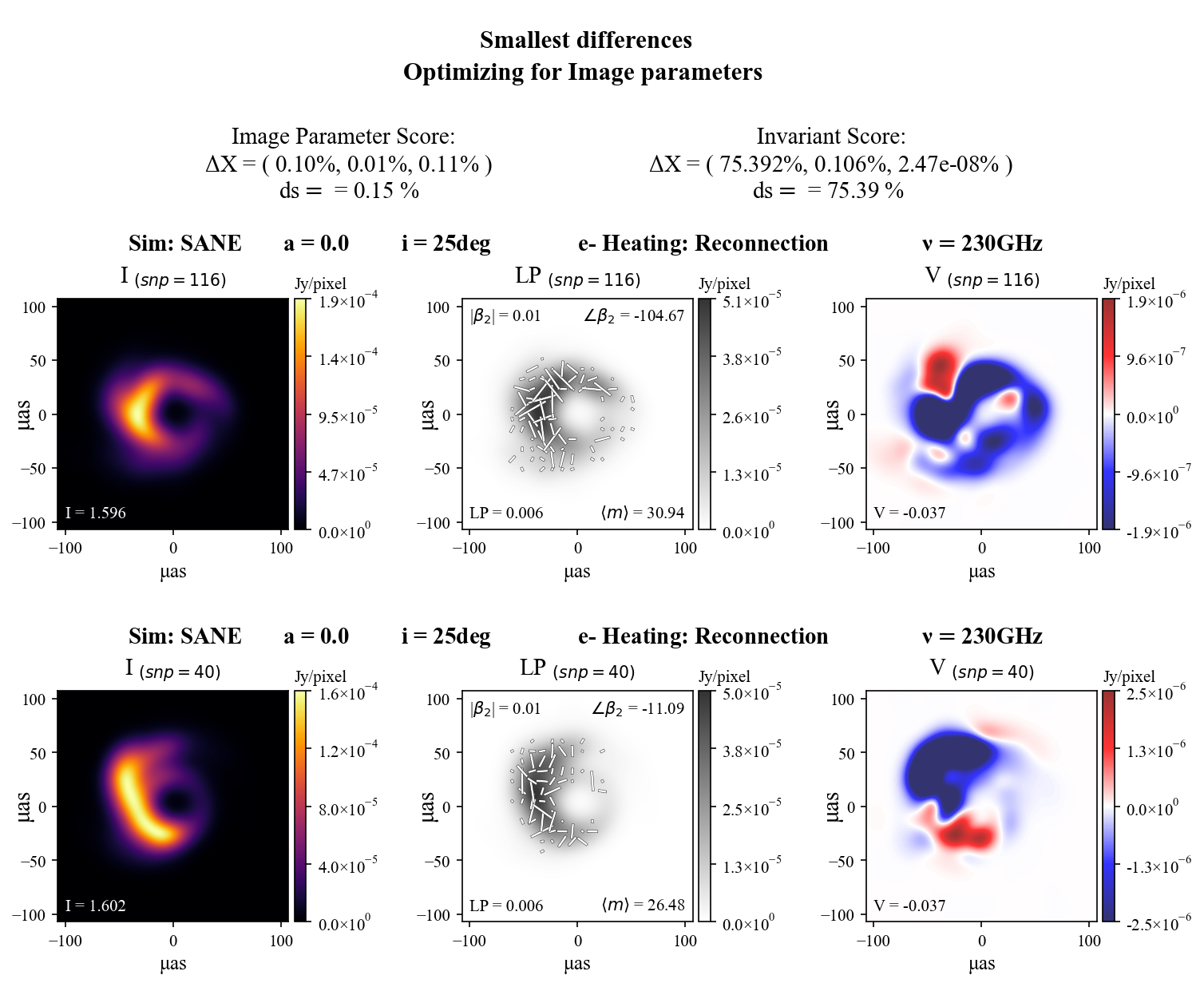}
\caption{Closest frames from two different models in our entire library according to criterion 1: the joint differences between image-integrated $I$, $LP$ and $V$, are the smallest. Both models are SANE RC, with zero spin. Each column shows, from left to right, $I,~LP,~V$. The value of each image-integrated quantity is shown at the bottom of the panel. Linear polarization shows as well the resolved polarization fraction $\langle m \rangle$ and $\beta_2$ coefficients (see main text for details). The selected ``closest'' frames from each model are indicated at the top with the label "snp". Two scores are shown at the top of the figure. Left: $ds$ (Eq.~\ref{eq:ds}) following $\Delta X$ from (Eq.~\ref{eq:deltaX_param}). Right: $ds$ following $\Delta X$ from (Eq.~\ref{eq:deltaX_inv}). The same fixed frames are used for both calculations. 
Since the scoring algorithm has been optimized for image-integrated quantities, this frame combination is chosen by \textit{minimizing} $ds$ from $\Delta X$ given by Eq.~\ref{eq:deltaX_param}. It can be seen that while the image-integrated values between the frames are very similar, the overall morphology of these images is rather different and is reflected in the large value of $ds$ calculated from the HF invariants that characterize these images.}
\label{fig:closest_imparams}
\end{figure*}

\begin{figure*}
\centering
\includegraphics[trim={0cm 0cm 0cm 0.5cm},clip=True,width=1.8\columnwidth]{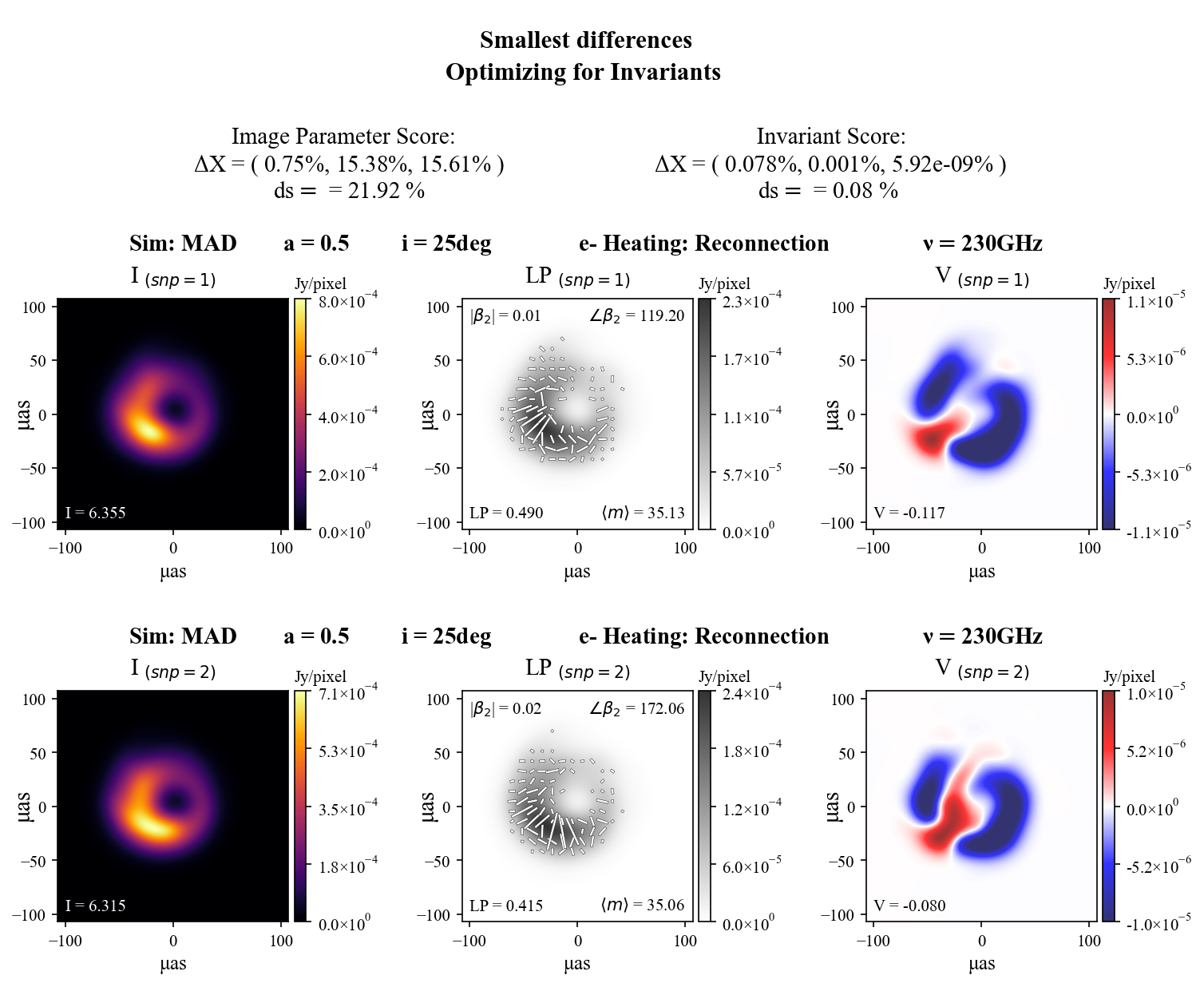}
\vspace{-0.3cm}
\caption{Similar to Fig.~\ref{fig:closest_imparams} but optimized for criterion 2. The frames and models are chosen out of all the possible combinations in our entire library (except for frame self-comparison), by \textit{minimizing} $ds$ from $\Delta X$ given by Eq.~\ref{eq:deltaX_inv}, so that the joint differences between the invariants that characterize the $I$, $LP$ and $V$ images is the smallest. Both frames come from the same model: MAD with spin $a=0.5$ and RC as a heating mechanism. It can be seen that the distance score based on the image-integrated values is larger compared to that in Fig.~\ref{fig:closest_imparams}. The overall morphology of the $I$, $LP$ and $V$ images is much more similar, including the spatial distribution of linear polarization vectors (white ticks in foreground) in the $LP$ panel, which also display a much more ordered configuration. }
\label{fig:closest_huinvs}
\end{figure*}

\begin{figure*}
\centering
\includegraphics[trim={0cm 0cm 0cm 0.5cm},clip=True,width=1.8\columnwidth]{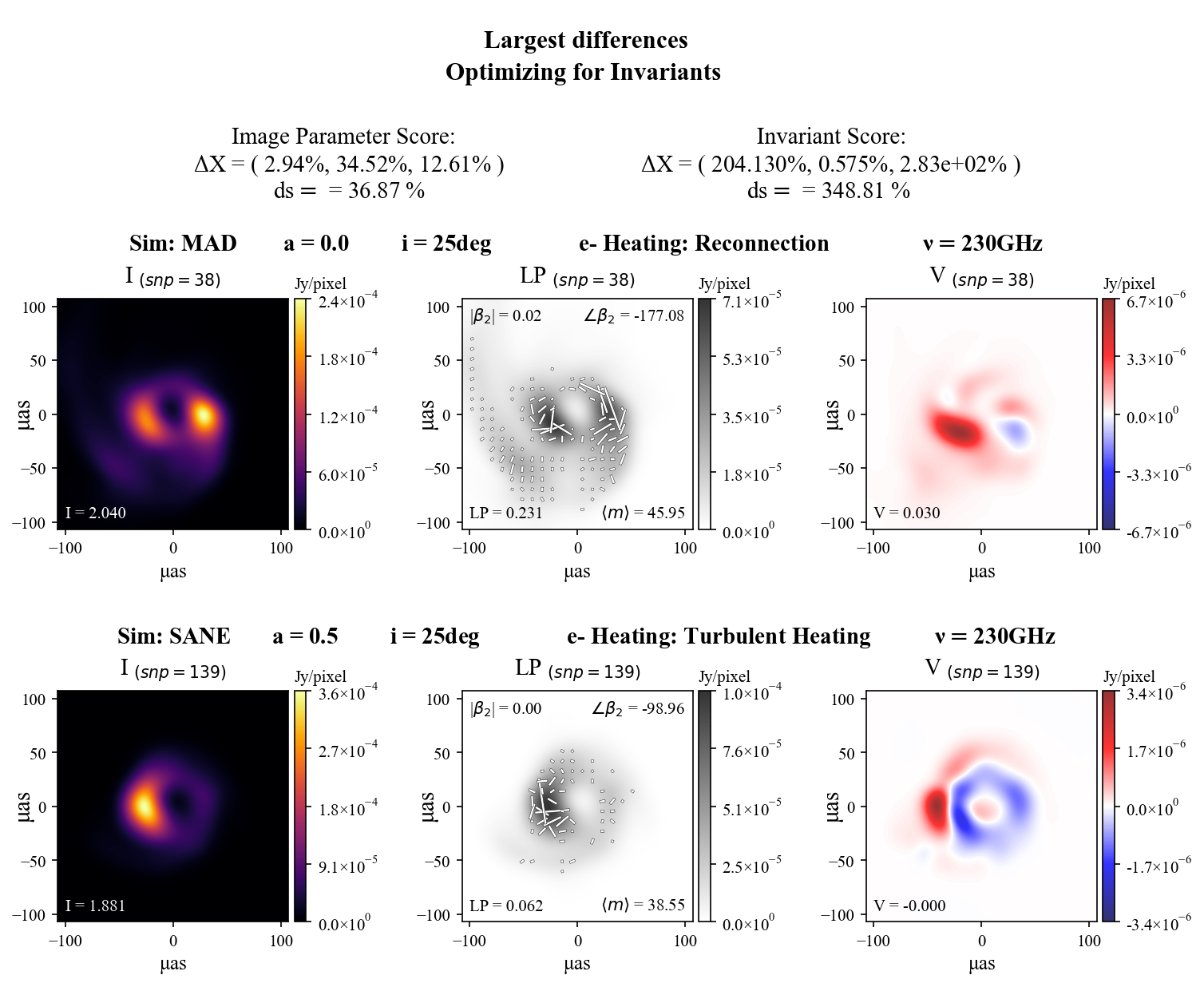}
\caption{Similar to Fig.~\ref{fig:closest_huinvs} but maximizing the distance score obtained according to criterion 2, i.e. the joint differences between the invariants that characterize the $I$, $LP$ and $V$ images are the largest of all combinations. In this case, the most different frames in our library correspond to those from a MAD and SANE, non-zero spin with magnetic reconnection as the electron heating mechanism. It is clear that the images exhibit different morphologies and spatial distribution of polarization vectors. The distance score from invariants is also much larger than that obtained from image parameters.  }
\label{fig:most_different_huinvs}
\end{figure*}

A few quantities that contain image spatial information have been considered in past works to differentiate between models and real black hole images. Those include second-order image moments to measure image sizes or asymmetries \citep[e.g.][]{EHTPaperV,EHTSgraPaperV}, or spatially resolved linear polarization fractions and maps \citep[e.g.][]{palumbo2020,EHTPaperVII,EHTPaperVIII}, some only interpretable at low inclinations. Also, when scoring a model, quantities such as Stokes $I$, $LP$ and Stokes $V$ are often considered independently of each other.

In this work, we aim to improve current scoring methods in two ways. The first improvement is to use image moment invariants, which encode structural information of the model images, uniformly for total intensity and polarimetric images. The second improvement is to combine the value of $I$, $LP$ and $V$ invariants into one when scoring a model. 

Our scoring procedure is based on calculating a ``distance score'', $ds$, between two model frames (or between a model and an observed image):

\begin{equation}
\label{eq:ds}
ds = \lVert ~\Delta X~ \rVert = \sqrt{ \sum_i{ \Delta X_i^2 } } ;
\end{equation}

\noindent where $\Delta X$ is a vector made up of the differences between multiple quantities according to different criteria:

\begin{enumerate}[wide, labelwidth=!,itemindent=!,labelindent=0pt, leftmargin=0em, label=(\arabic*), itemsep=1.2cm, parsep=0pt]
    {\item Image-integrated parameters:} \\
    \begin{equation} 
    \label{eq:deltaX_param}
    \Delta X~ =~(~\Delta I_{mn},~\Delta LP_{mn},~\Delta V_{mn}~), 
    \end{equation}
    \noindent where each entry is the percentile difference of a polarized quantity between the frames $m$ and $n$ of their respective model:
    
    \begin{equation} 
    \label{eq:percentile_diff}
    \Delta Y_{mn} = \frac{|Y_m - Y_n|}{\textrm{max}(Y)-\textrm{min}(Y)} \times 100 \% ; ~~Y~=~(~I,~LP,~V~)
    \end{equation}

    \noindent where $\textrm{max}(Y)$ and $\textrm{min}(Y)$ are the maximum and minimum values of $Y$ in the entire image population. 
    \newline
    {\item Invariants:} \\
    \begin{equation}
    \label{eq:deltaX_inv}
    \Delta X~=~(~\lVert ~\Delta (HF)_{I_{mn}}~ \rVert,~\lVert ~\Delta (HF)_{LP_{mn}}~ \rVert,~\lVert ~\Delta (HF)_{V_{mn}}~ \rVert~)     
    \end{equation}    
    where $\Delta (HF)_{Y_{mn}}=(~\Delta (HF_0)_{Y_{mn}},~...~,~\Delta (HF_7)_{Y_{mn}}~)~;~Y=(~I,LP,V~)$; is a vector made up of the percentile differences (Eq.~\ref{eq:percentile_diff}) between the HF invariants given in Eq.~\ref{eq:hu_invariants}.
\end{enumerate}

Following this procedure, we define the ``closest'' and ``farthest" models as those with the smallest and largest $ds$ between them:
\begin{equation}
\label{eq:min_ds}
\textrm{min}(~ds~) = \textrm{min}(~\lVert ~\Delta X~ \rVert~)~; \\
\textrm{max}(~ds~) = \textrm{max}(~\lVert ~\Delta X~ \rVert~).
\end{equation}

We note that in this work the weight of each component of $\Delta X$ is the same, as is each contribution from the different invariants (in case of criterion 2). This can be easily adapted to accommodate different observational uncertainties coming from measurements.

We have applied this scoring procedure to every two-frame combination ($150\times150$ frames), with the exception of self-frame comparison (150), from every two models in our library (78 combinations). In total, we find the closest and farthest frames from two models in our library from a sample of 1800 images and 1,743,300 possible combinations. 
Figures~\ref{fig:closest_imparams} and \ref{fig:closest_huinvs} show examples of $I,~LP,~V$ images of the \textit{closest} models in our library according to either criterion 1 (optimizing for image parameters) or 2 (optimizing for invariants). 

In the figures each column shows, from left to right, the $I,~LP,~V$ images of either model. The specific closest frame selected from a model is specified at the top of the panels with the label "snp". The value of each image-integrated quantity is shown at the bottom. In the case of linear polarization, three more quantities which encapsulate spatial information of the polarization are shown: the resolved linear polarization fraction $\langle m \rangle$ and the $\beta_2$ coefficients used to characterize the spatial distribution of polarization vectors at low inclinations \citep{palumbo2020}. Two $ds$ scores are shown at the top of the figure. On the left, $ds$ is calculated using $\Delta X$ from Eq.~(\ref{eq:deltaX_param}) while on the right, $\Delta X$ is defined by Eq.~(\ref{eq:deltaX_inv}). The same frames are used for both calculations.\footnote{We note that the scoring we have suggested for IMI is not the same that is currently used for image-integrated quantities so it is an unfair comparison. These other metrics that have been used in model comparison have their merits and give reasonable results. The comparison we make to HF invariants is not meant to discredit them, but rather to show the effectiveness of the IMI and to consider them as a supplementary approach. }

In the case of Fig.~\ref{fig:closest_imparams}, the closest frames come from two SANE RC models, both with spin zero, while for Fig.~\ref{fig:closest_huinvs}, the closest frames come from the same MAD RC with $a=0.5$ model.

It is evident that very different kinds of models can generate frames with very similar image-integrated quantities (Fig.~\ref{fig:closest_imparams}) but with images that look quite distinct between each other. This causes a very large difference with respect to the value of the invariants. 
When taking into consideration the structure and morphology of the image (Fig.~\ref{fig:closest_huinvs}), similar images can be found between distinct models. However, the differences using invariants are comparatively larger than when considering criterion 1 based on image-integrated quantities, pointing to a higher sensitivity and discerning power of the former. Both frames come from the same model: MAD with spin $a=0.5$ and RC as a heating mechanism. Since the frames are highly correlated (see Section~\ref{sec:time}), the algorithm naturally chooses consecutive frames. Even so, the invariants pick up on the fact it is not the same image and is still larger than that for image parameters, pointing to a higher sensitivity and discerning power of these quantities.

It is interesting to note that in Fig.~\ref{fig:closest_huinvs}, the polarization maps (white ticks in foreground of LP panels), which indicate the spatial distribution of the linear polarization vectors, show very similar configurations between the frames. The value of the resolved linear polarization fraction $\langle m \rangle$ differs by less than $2\%$. The $\beta_2$ coefficients, on the other hand, show more difference in the consecutive frames: the magnitude of the $\beta_2$ coefficient changes by $50\%$ while the angle coefficient has a much larger difference between the values.   

In Fig.~\ref{fig:most_different_huinvs} we show as well the two frames that are the most different in our library according to their invariants. This is achieved by \textit{maximizing} $ds$ obtained from criterion 2. It can be seen that the structural differences between the images are far more evident, which is reflected in the large value of the distance score for invariants.

\section{Time evolution of invariants}
\label{sec:time}

\begin{figure*}
\centering
\includegraphics[trim={0cm 0cm 0 0cm},clip=True,width=1.8\columnwidth]{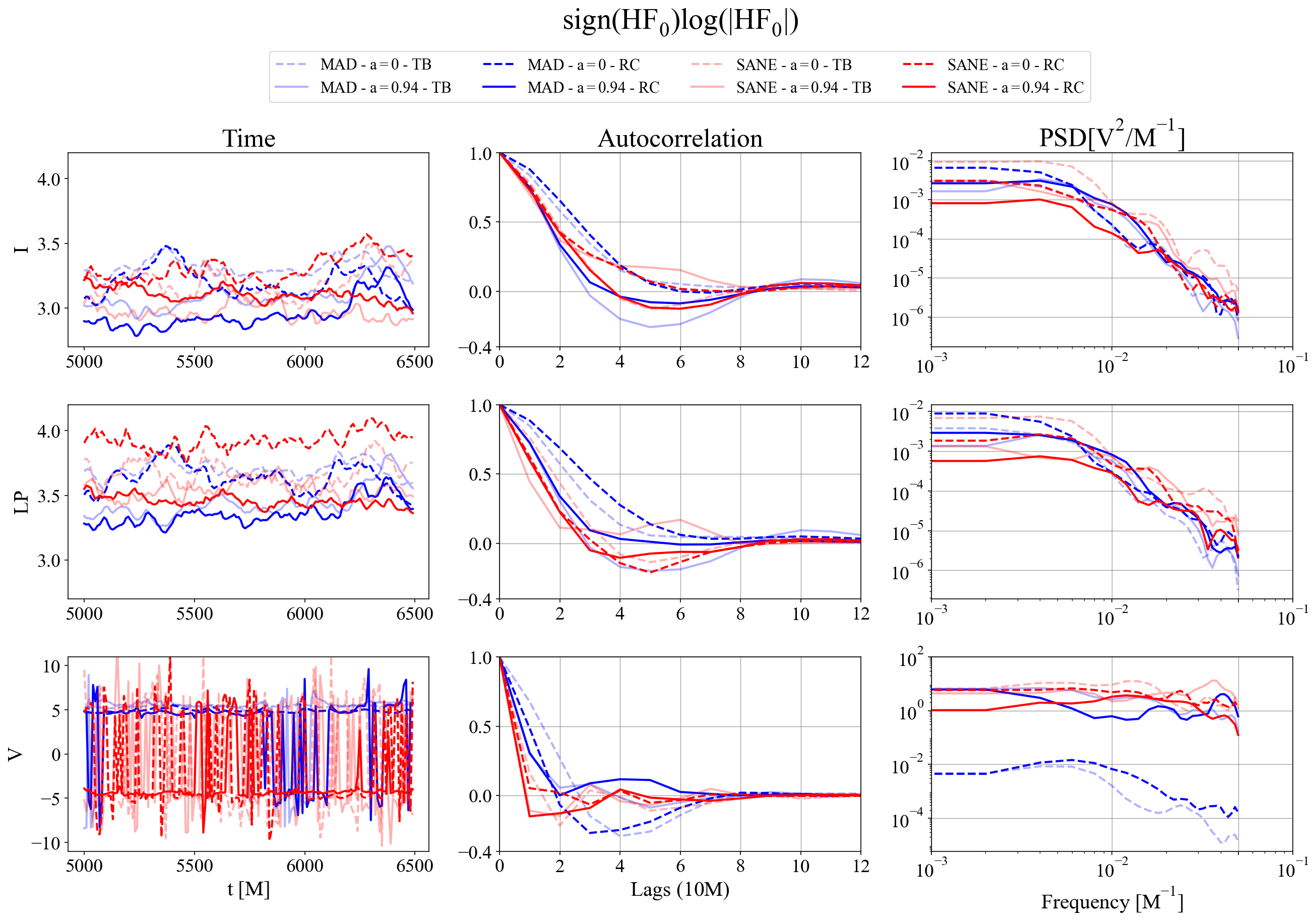}
\caption{Left column: time evolution of the $g(HF_0)$ value for $I$, $LP$ and $V$ for all simulations. Middle column: the autocorrelation of the time series for the first 12 snapshots (120 M). 
Right column: the power spectrum density of the time series.} 
\label{fig:evolution}
\end{figure*}

Studying the time evolution of invariants is also of great interest, since it provides a powerful new tool to compare models and data from multiple epoch observations, where structural changes in the source's image may be present. Specially after the results of \cite{EHTSgraPaperIV,EHTSgraPaperV}, where the quick variability of \sgra complicates the imaging process, but provides an opportunity to test theoretical models. With more data to come from more observing campaigns with the EHT, the analysis and modelling of these sources in the time domain are growingly becoming more important.  

In Fig.~\ref{fig:evolution} we show a variety of different quantities that describe the time properties of $g(HF_0)$ for $I$ (top row), $LP$ (middle row) and $V$  (bottom row) for all models. The first column shows the time evolution of all three observed quantities for the whole duration of our simulations. There appears to be no particularly characteristic feature in the curves, though it is interesting to mention that the behavior for Stokes V is variable over a wider range of values, presumably due to a switch in ``polarity'' of the quantity with time. 

In the second and third columns we present, respectively, the autocorrelation and the power spectral density (PSD) of the time series. 
Both of these calculations were done using \texttt{scipy.Welch} \citep{Welch,2020SciPy-NMeth} with an overlapping window of 50 snapshots. This serves as a smoothing kernel and allows us to observe more clearly a trend in the data. 

In the case of the autocorrelation function, we show the first 120 M (limit set by the overlapping window of 50 snapshots). We observe that the frames become de-correlated on a timescale of 20-30 M, as indicated by the drop to zero of the curves. Stokes V appears to drop steeper than $I$ and $LP$. 

As previously mentioned, in the last column we plot the PSD of the time evolution. This is useful to discover any recurring behavior in the data, for example a peak at $F=0.01~M^{-1}$ would be indicative of periodic behavior every $100~M$. The other, and main reason, is to identify the existence of a general trend which could be, for example, in the form of a power-law.   
Even though the spacing between the frames in our library is relatively big (10 M) and the duration of the models is short (1500 M), which limits the frequency space that we are able to probe, we find an interesting behavior. For $I$ and $LP$, the spectrum can be described as a plateau at low frequencies, with a power-law drop-off at higher ones. Interestingly, this is not observed for Stokes V. 
This points towards the presence of red noise in the data, reminiscent of what has been found for image-integrated total intensity from a large set of EHT GRMHD simulations \citep{Georgiev_2022}.  

We have calculated these quantities for all HF invariants, but decided to show only those for $HF_0$ in the interest of simplicity. A generic observation is that higher order invariants de-correlate immediately.

Such an extensive analysis in the time domain of the invariants from all observables ($I$, $LP$ and $V$), for all the models is a powerful tool.  These new variability constraints could be combined and used as a prescription for aiding imaging algorithms \citep[e.g.][]{Broderick22_variability}, or parameter estimation pipelines \citep[e.g.][]{yfantis2023}.

\section{Discussion} 
\label{sec:discussion}

In this work we have explored how geometric IMI can be used as a new method for model discrimination of accretion onto BH. We have used a library of polarized images calculated from a variety of models from GRMHD simulations and calculated their IMI. Since IMI encapsulate structural changes of the images (though it is still unclear what exact image property is measured by each one; Appendix~\ref{appdx:min_max_diff}), we have shown that they can be highly sensitive to different physical effects present in the system (e.g. magnetization, spin of the black hole and electron heating mechanism; see Section~\ref{sec:results1}). These distributions could be used to identify the probability that a given (calculated or measured) image belongs to a population with certain parameters. Given the modest size of our library, we leave this for future work.

Current model scoring methods consider a variety of properties of the images they produce and compare to the corresponding observables. Each quantity, however, is often calculated with different approaches and a final model score, which could give sensible results, is considered independently. 
We have proposed a new scoring method that is based on IMI and is not only applicable to total intensity and polarimetric images uniformly, but also combines the value of $I,~LP$ and $V$ invariants into one (see Section~\ref{sec:results2}). We do not attempt to quantify EVPA images because they may be a subject to external Faraday rotation caused by material far out in regions outside the domain of the models. An application to EHT data is left for future work. 

Given the powerful properties of the IMI, if it were possible to disentangle the EHT beam resolution with the “true” image underneath it would be possible to conduct mass measurements with the observations and mass-agnostic model comparison. Unfortunately it is not clear how to do it exactly. Perhaps there exists an extension of the IMI basis to Fourier space, where the observations are made. Whether this is possible while keeping the properties of the IMI intact, we leave for future work. In addition, it is worth to mention that future EHT arrays and observations (either at 345 GHz or space VLBI) will have better resolution and the blurring will be smaller, which is an exciting prospect to look forward to.

Still, this technique as is has an advantage over directly model fitting the visibilities of VLBI data, since the images are already an excellently calibrated data set and there is no need for further calibration. 
Moreover, our new procedure is versatile with the type of image morphologies that are considered. This means that it is not limited to ring-like structures and could be applied to other kinds of images with more extended or elongated features such as jets, which have been observed at longer wavelength VLBI observations \citep[e.g. 22, 43, and 86~GHz][]{kravchenko2020,park2021}. 
We explore this in Appendix~\ref{appdx:jet_images}, where we have applied our algorithm to images with prominent jets in our model library. We find that our scoring method works as well for these non-ring images, as expected. 
We have also applied the scoring method to an extended sample of images that go beyond accretion onto black holes (Appendix~\ref{appdx:extended_library}). We find that the algorithm successfully judges and finds the two black hole images in the collection as those with the smallest distance between them.

In this analysis we only considered models at low viewing angles, but the application to other viewing angles is straight forward. We observed that, as a function of inclination, the overall shape of the distributions remains very similar, only at high inclinations are the changes slightly more evident. 

We have also explored the effect that blurring has on the HF distributions. 
This is of particular interest since the EHT will observe at 345 GHz in the future, for which the resolution will be better by $40\%$. The general behavior we observe is that the differences between the invariants of different images become smaller and converge to zero as the size of the blurring kernel increases.  
For the particular improved resolution the EHT at 345 GHz, the changes to the invariant distributions are very small and not particularly evident when shown in a logarithmic scale. As a consequence, the invariant distributions will remain very similar between 230~GHz and 345~GHz.

Lastly, we have studied the time-dependent behavior of the invariants. We show that the models de-correlate at scales of 20-30~M. The power spectral distributions of the $I$ and $LP$ invariants show a plateau at low frequencies that falls like a power-law at high frequencies. This is not observed for Stokes V. This power-law trend seems to be consistent with other findings by \citet{Georgiev_2022} regarding light-curve variability in BH accretion GRMHD simulations. 

In this work we have shown that IMI are a promising new approach for characterizing the nature of astrophysical systems and will certainly prove useful when learning about the intricacies of magnetized accretion onto massive black holes.


\section*{Acknowledgements}
We thank  H. Olivares, J. Vos, M. Bauböck, G. Wong, C. Brinkerink and the anonymous referee for useful and helpful discussions. We acknowledge support from Dutch Research Council (NWO), grant no. OCENW.KLEIN.113.
This publication is part of the project the Dutch Black Hole Consortium (with project number NWA 1292.19.202) of the research programme the National Science Agenda which is financed by the Dutch Research Council (NWO).

\section*{Data Availability}

The simulated images used here will be shared on reasonable request to the corresponding author.



\bibliographystyle{mnras}
\bibliography{mnras_template} 



\appendix

\section{Moment invariants verification and properties}
\label{appdx:verification}

Here we demonstrate that our script for calculating IMHI is correct. We show the invariance of IMHI with respect to translation, rotation and scaling. Without loss of generality, we use one of the frames of a non-zero spin, MAD reconnection model. 

\begin{figure*}
\centering
\includegraphics[trim={0cm 0cm 0cm 0cm},clip=True,width=2.1\columnwidth]{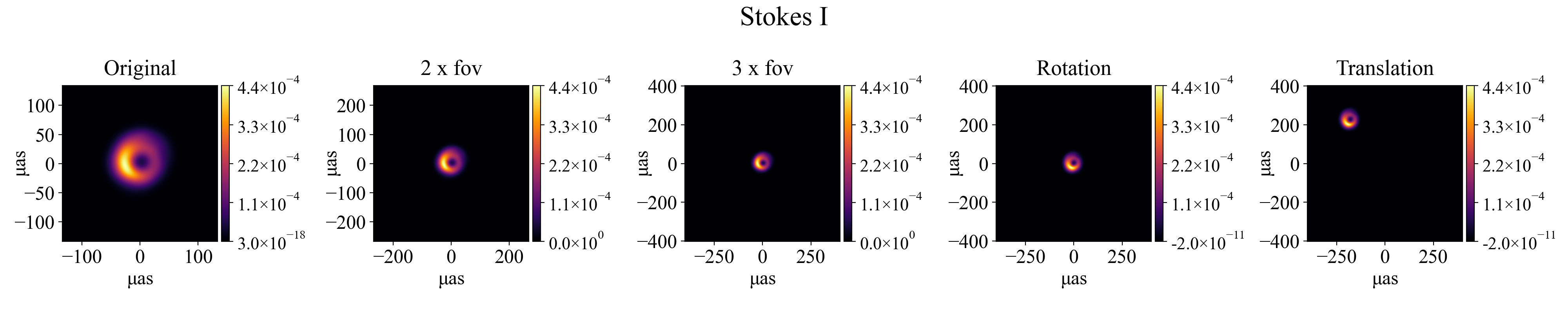}
\vspace{-0.4cm}
\caption{Stokes $I$ of a randomly selected frame from our library. Each column shows the image being modified by a sequential transformation. These are, in order: original, scaled fov to twice the original, increased fov by three times, rotation of $45\deg$, translation of image centre from $(0,0)~\mu as$ to $(-160,200)~\mu as$.  }
\label{fig:inv_verification_I}
\end{figure*}

Each column of Fig.~\ref{fig:inv_verification_I}, shows the Stokes $I$ image under the following (sequential) transformations: original, scaling of field-of-view (fov) by a factor of two, scaling of fov by factor of three, rotation (randomly selected value of $45\deg$) and translation, where centre of the image was shifted, randomly, from $(0,0)~\mu as$ to $(-160,200)~\mu as$.

The values of the percentile differences of the HF invariants for each transformed image in Fig.~\ref{fig:inv_verification_I} and the original are shown in Table~\ref{table:inv_verification_huI}. 

\begin{figure*}
    \centering
    \includegraphics[trim={0cm 0cm 0cm 0cm},clip=True,width=2.1\columnwidth]{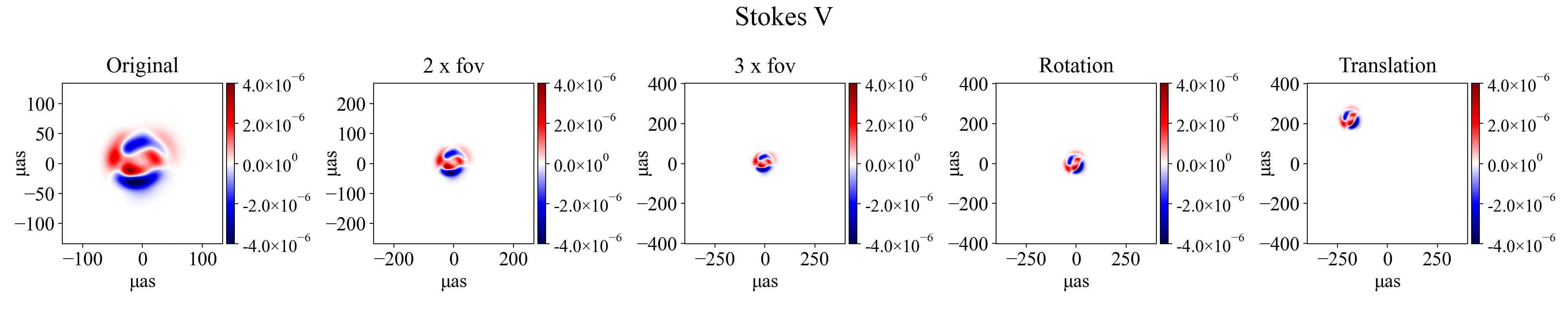}
    \vspace{-0.4cm}  
    \caption{Same as Fig.~\ref{fig:inv_verification_I} but for Stokes V.}
    \label{fig:inv_verification_V}
\end{figure*}

\begin{figure*}
\centering
\begin{subfigure}[t]{0.48\textwidth}
    \centering
    \includegraphics[trim={0cm 0cm 0cm 0cm},clip=True,width=0.95\columnwidth]{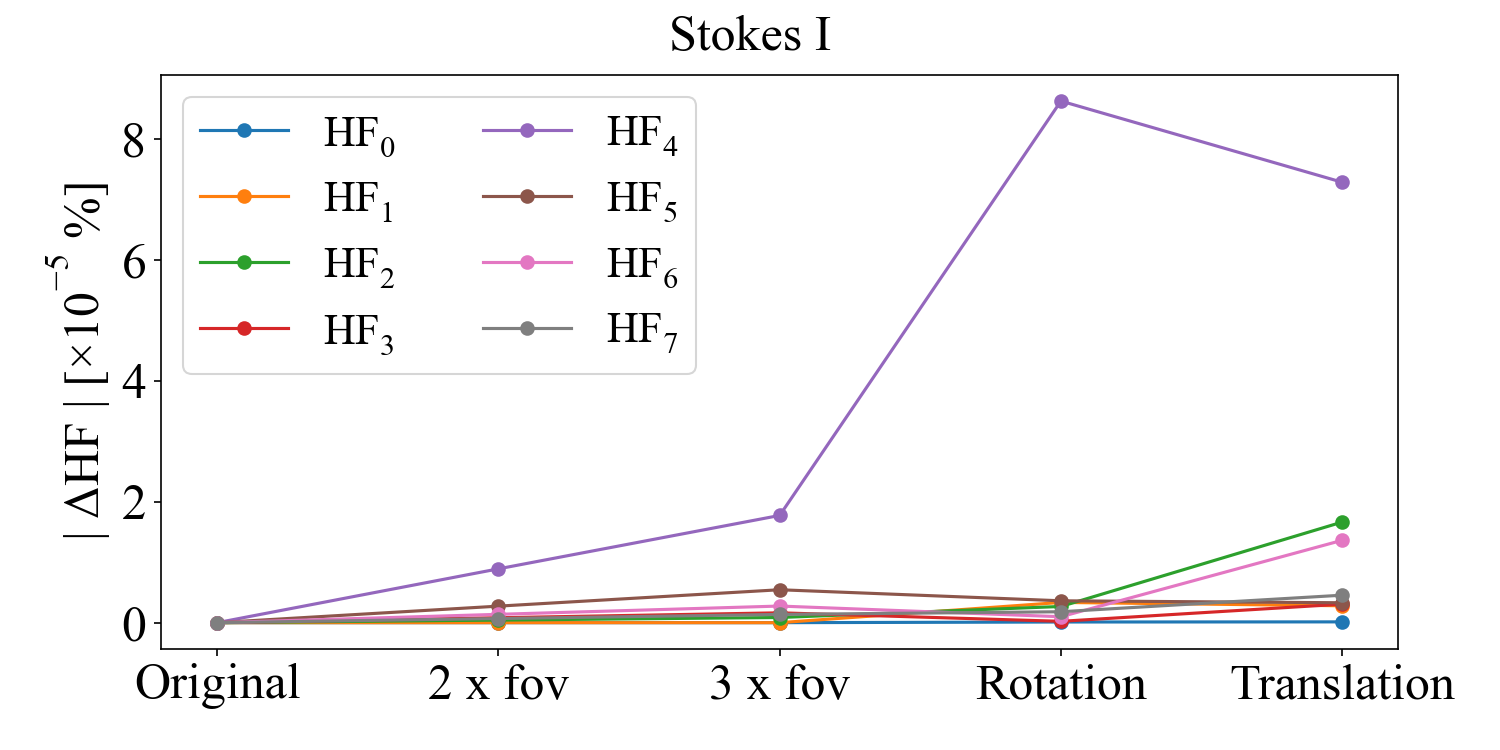}
    \caption{ }
    \label{fig:inv_verification_huI}
\end{subfigure}
\begin{subfigure}[t]{0.48\textwidth}
    \centering
    \includegraphics[trim={0cm 0cm 0cm 0cm},clip=True,width=0.95\columnwidth]{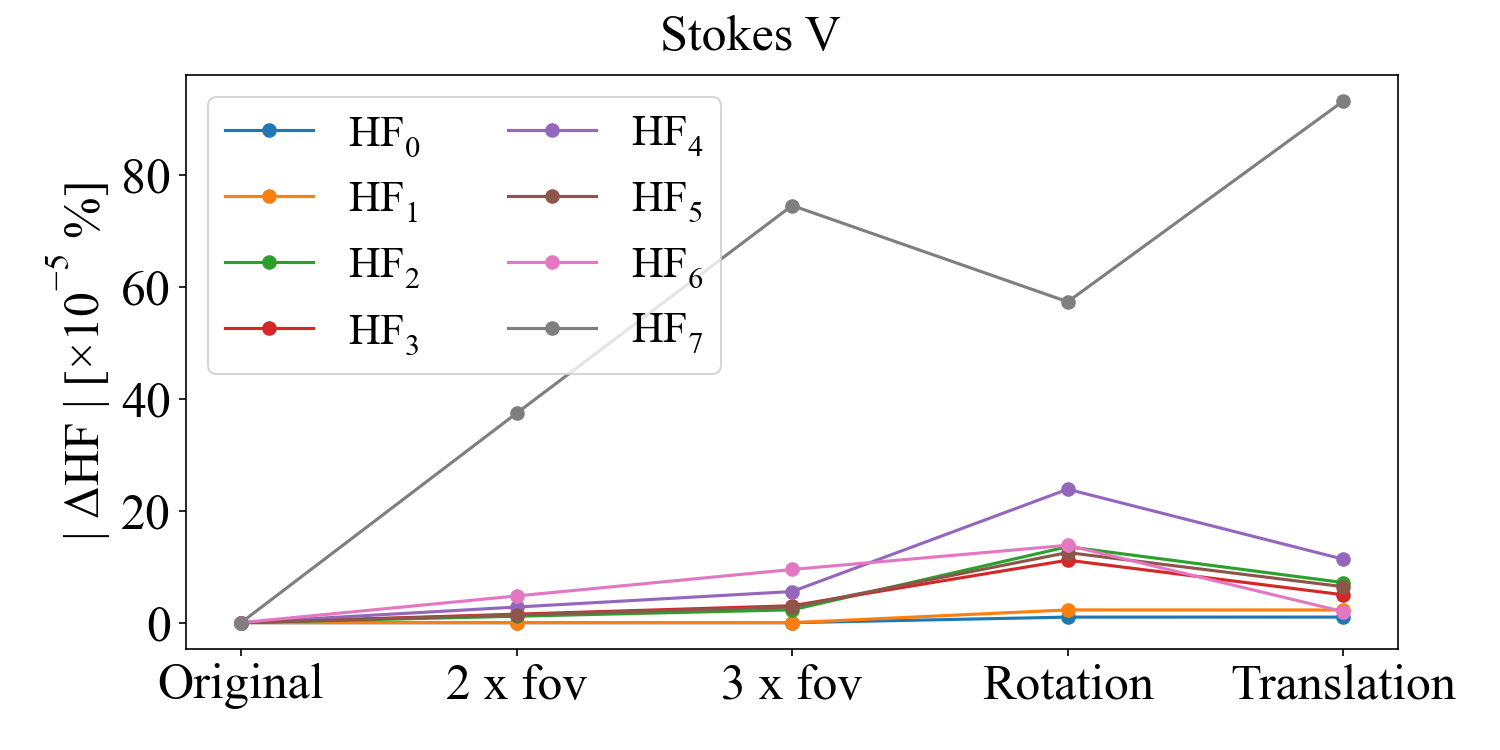}
    \caption{ }
    \label{fig:inv_verification_huV}
    \end{subfigure}
\caption{Percentile differences of HF invariants between the original and each transformed image from (a) Fig.~\ref{fig:inv_verification_I} and (b) Fig.~\ref{fig:inv_verification_V}.}
\label{fig:inv_verification}    
\end{figure*}

\begin{table*}
	\centering
	\caption{Percentile differences of the HF invariants between each transformed image in Fig.~\ref{fig:inv_verification_I} and the original.}
	\label{table:inv_verification_huI}
	\begin{tabular}{lcccccccc} 
		\hline
		        & |$\Delta HF_0$| [\%] & |$\Delta HF_1$| [\%] & |$\Delta HF_2$| [\%] & |$\Delta HF_3$| [\%] & |$\Delta HF_4$| [\%] & |$\Delta HF_5$| [\%] & |$\Delta HF_6$| [\%] & |$\Delta HF_7$| [\%]\\
		\hline
		Original - 2 x fov  & 5.62e-14 & 2.96e-13 & 4.29e-7 & 8.18e-7 & 8.90e-6 & 2.73e-6 & 1.38e-6 & 6.84e-7\\
		Original - 3 x fov  & 9.83e-14 & 9.87e-14 & 8.54e-7 & 1.62e-6 & 1.77e-5 & 5.44e-6 & 2.75e-6 & 1.36e-6\\
		Original - Rotation & 1.10e-7 & 3.35e-6 & 2.68e-6 & 2.23e-7 & 8.63e-5 & 3.62e-6 & 9.81e-7 & 1.81e-6\\
		Original - Translation & 1.31e-7 & 2.79e-6 & 1.64e-5 & 3.07e-6 & 7.23e-5 & 3.29e-6 & 1.36e-5 & 4.55e-6\\
		\hline
	\end{tabular}
\end{table*}

Fig.~\ref{fig:inv_verification_huI} shows the percentile differences of the HF invariants between each transformed image from Fig.~\ref{fig:inv_verification_I} and the original. It can be seen that the values of the invariants of the transformed images remain very similar with respect to the original. The largest difference arises with rotation. This is due to numerical errors introduced by interpolated values to new ``in-between'' pixels of the image. However, even taking this into account, the largest differences are of the order of $10^{-5}\%$. 

Due to linear polarization images being always positive, we expect the behavior of the invariants to be similar to the one for Stokes $I$.

We explore the behavior of Stokes V images, where the value of a pixel value can be in the negative regime.
We use the Stokes $V$ image from the same frame as before and transform it in the same way as described for Stokes $I$.
Fig.~\ref{fig:inv_verification_V} shows the corresponding transformed images, while Fig.~\ref{fig:inv_verification_huV} shows the value of the percentile differences between the transformed Stokes $V$ images and the original. The values of these differences are presented in Table~\ref{table:inv_verification_huV}.

\begin{table*}
	\centering
	\caption{Percentile differences of the HF invariants for each transformed image in Fig.~\ref{fig:inv_verification_V} and the original.}
	\label{table:inv_verification_huV}
	\begin{tabular}{lcccccccc} 
		\hline
	        & |$\Delta HF_0$| [\%] & |$\Delta HF_1$| [\%] & |$\Delta HF_2$| [\%] & |$\Delta HF_3$| [\%] & |$\Delta HF_4$| [\%] & |$\Delta HF_5$| [\%] & |$\Delta HF_6$| [\%] & |$\Delta HF_7$| [\%]\\
		\hline
		Original - 2 x fov  & 5.41e-12 & 1.87e-11 & 1.15e-5 & 1.52e-2 & 2.79e-5 & 1.41e-5 & 4.77e-5 &3.74e-4\\
		Original - 3 x fov  & 1.61e-11 & 4.76e-11 & 2.29e-5 & 3.03e-5 & 5.57e-5 & 2.80e-5 & 9.51e-5 &7.46e-4\\
		Original - Rotation & 9.92e-6 & 2.28e-5 & 1.35e-4 & 1.18e-4 & 2.38e-4 & 1.25e-4 &   1.38e-4 & 5.73e-4\\
		Original - Translation & 9.96e-6 & 2.26e-5 & 7.17e-5 & 4.98e-5 & 1.13e-4 & 6.43e-5 & 1.97e-5 & 9.32e-4\\
		\hline
	\end{tabular}
\end{table*}

It can be seen that though the percentile differences are generally larger for Stokes $V$ compared to Stokes $I$, but the largest difference is of the order of $10^{-4}$, and so we conclude that the HF invariants for Stokes $V$ are indeed invariant.

\section{Moment invariant properties}
\label{appdx:properties}
\begin{figure*}
    \begin{subfigure}{1.0\textwidth}
    \centering
    {\normalsize Stokes~$I$}  \\ 
    \includegraphics[trim={0.0cm 1.5cm 0.0cm 2.5cm},clip=True,scale=0.3]{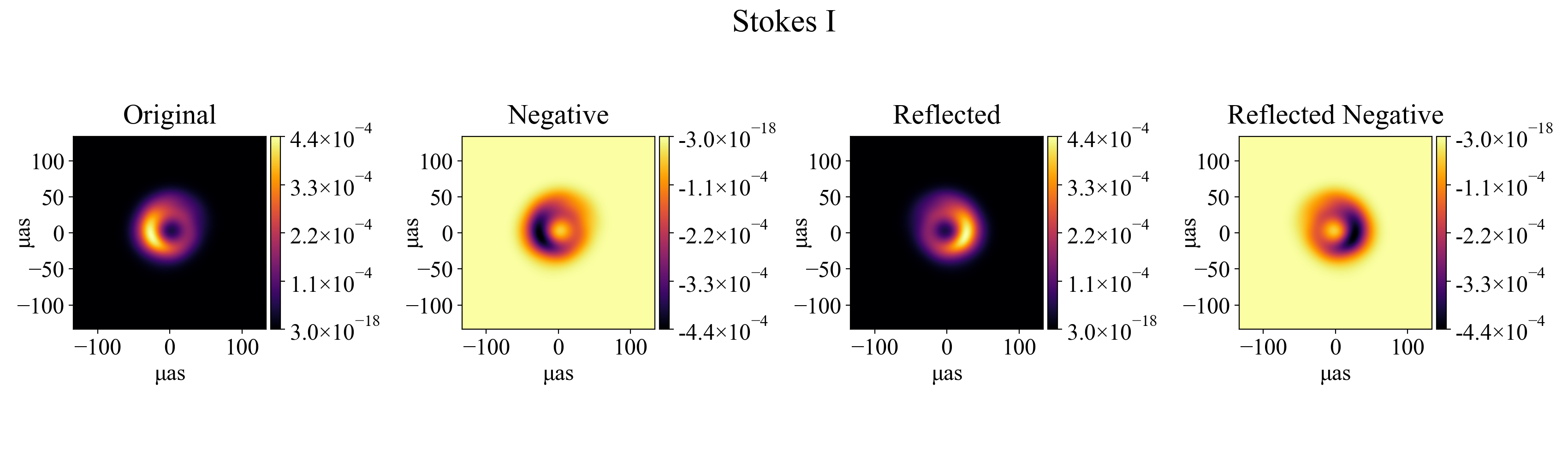}
    \includegraphics[trim={0cm 0cm 2.3cm 1cm},clip=True,scale=0.23]{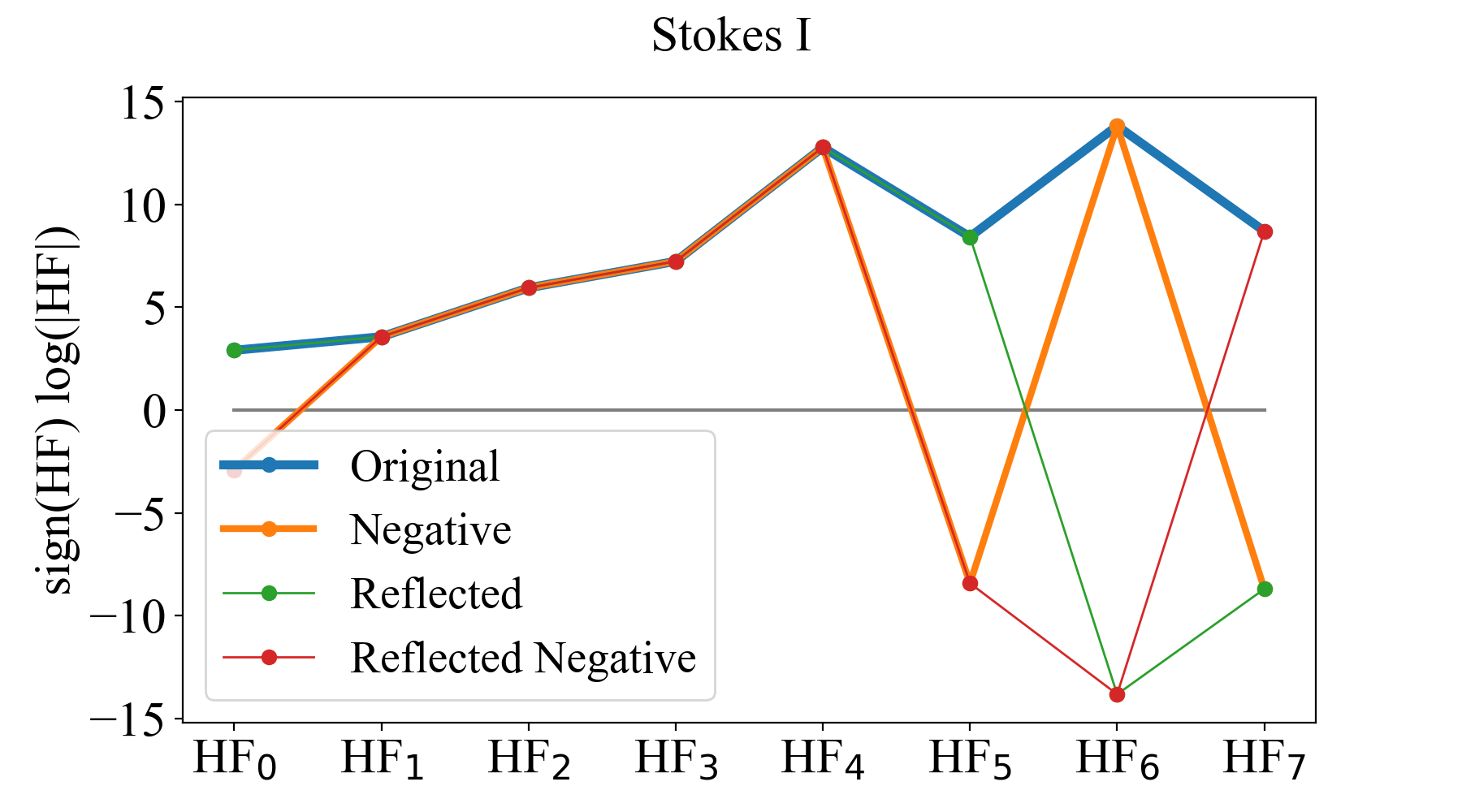}
    \end{subfigure}
    
    \begin{subfigure}{1.0\textwidth}
    \centering
    {\normalsize Linear Polarization} \\ 
    \includegraphics[trim={0.0cm 1.5cm 0.0cm 2.5cm},clip=True,scale=0.3]{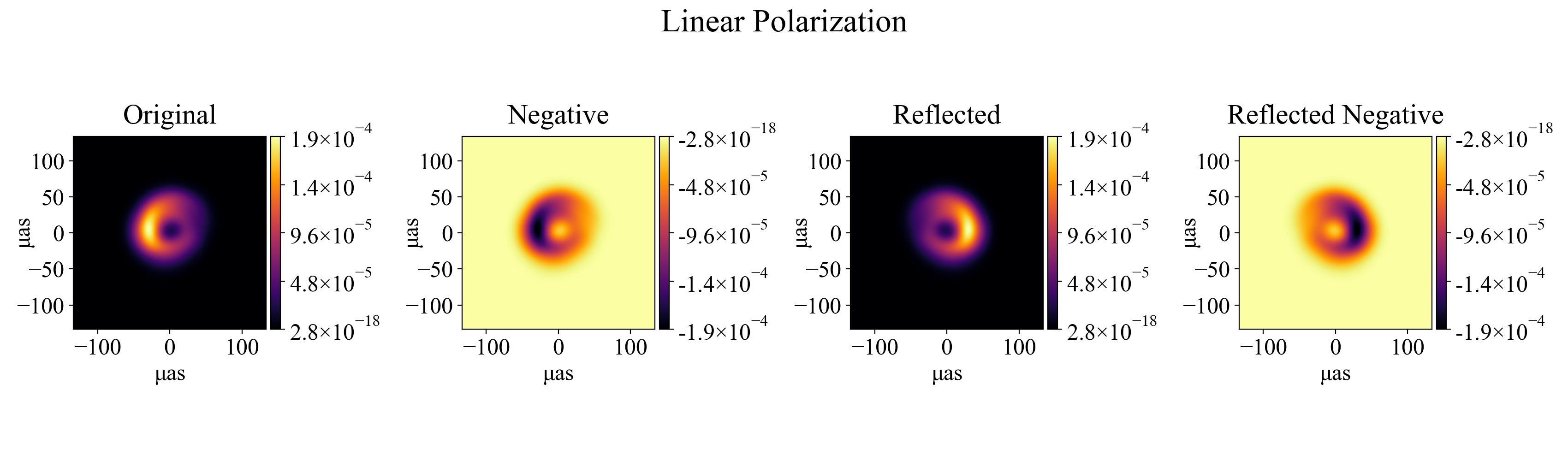}
    \includegraphics[trim={0cm 0cm 2.3cm 1cm},clip=True,scale=0.23]{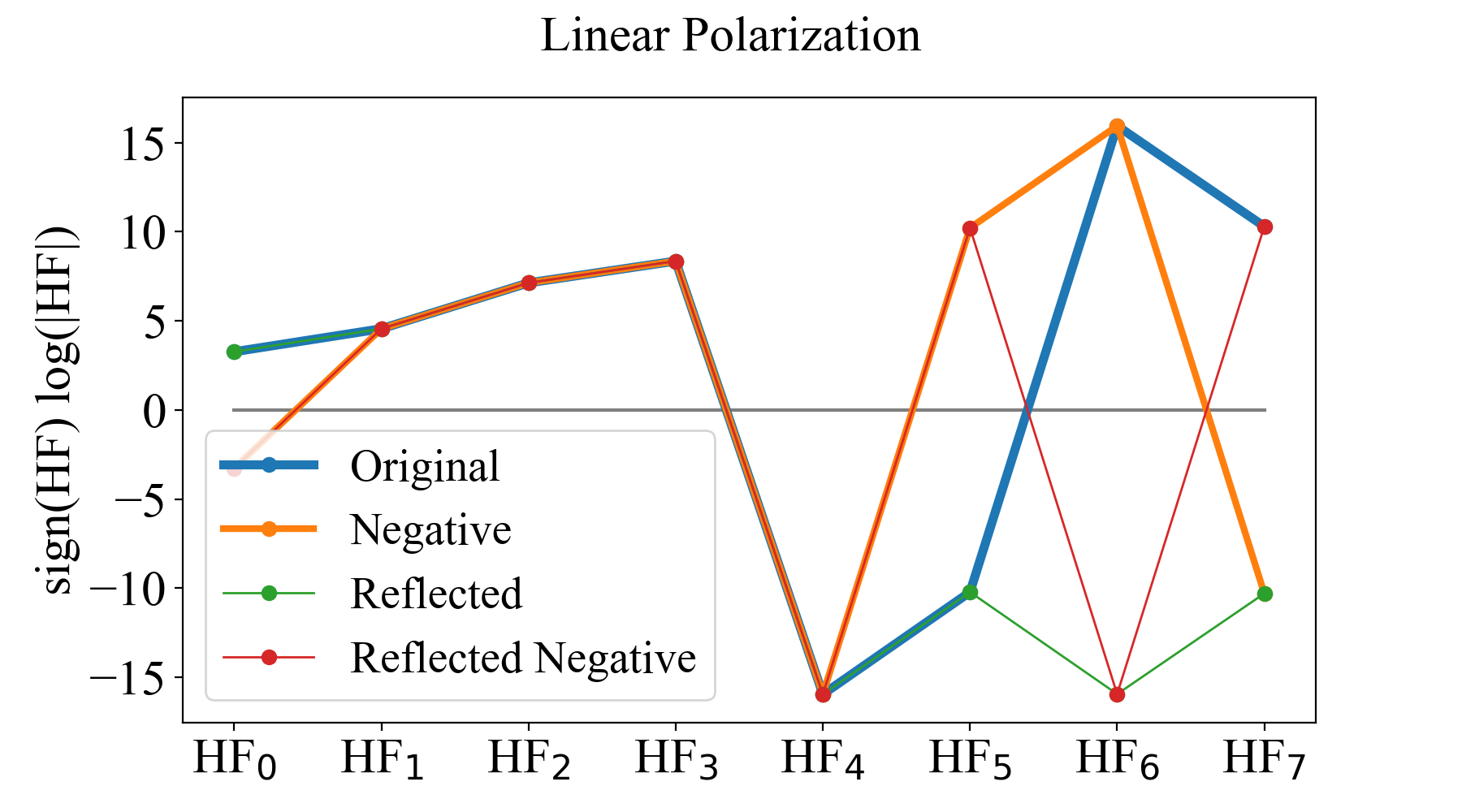}
    \end{subfigure}
    
    \begin{subfigure}{1.0\textwidth}
    \centering
    {\normalsize Stokes~$V$} \\ 
    \includegraphics[trim={0.0cm 1.5cm 0.0cm 2.5cm},clip=True,scale=0.3]{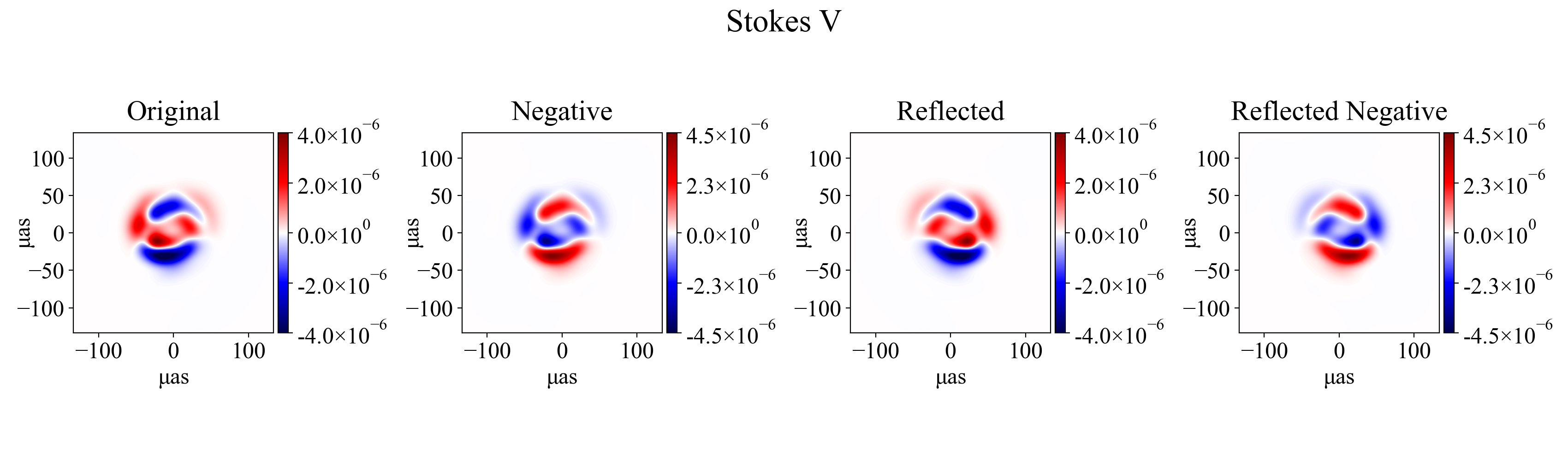}
    \includegraphics[trim={0cm 0cm 2.3cm 1cm},clip=True,scale=0.23]{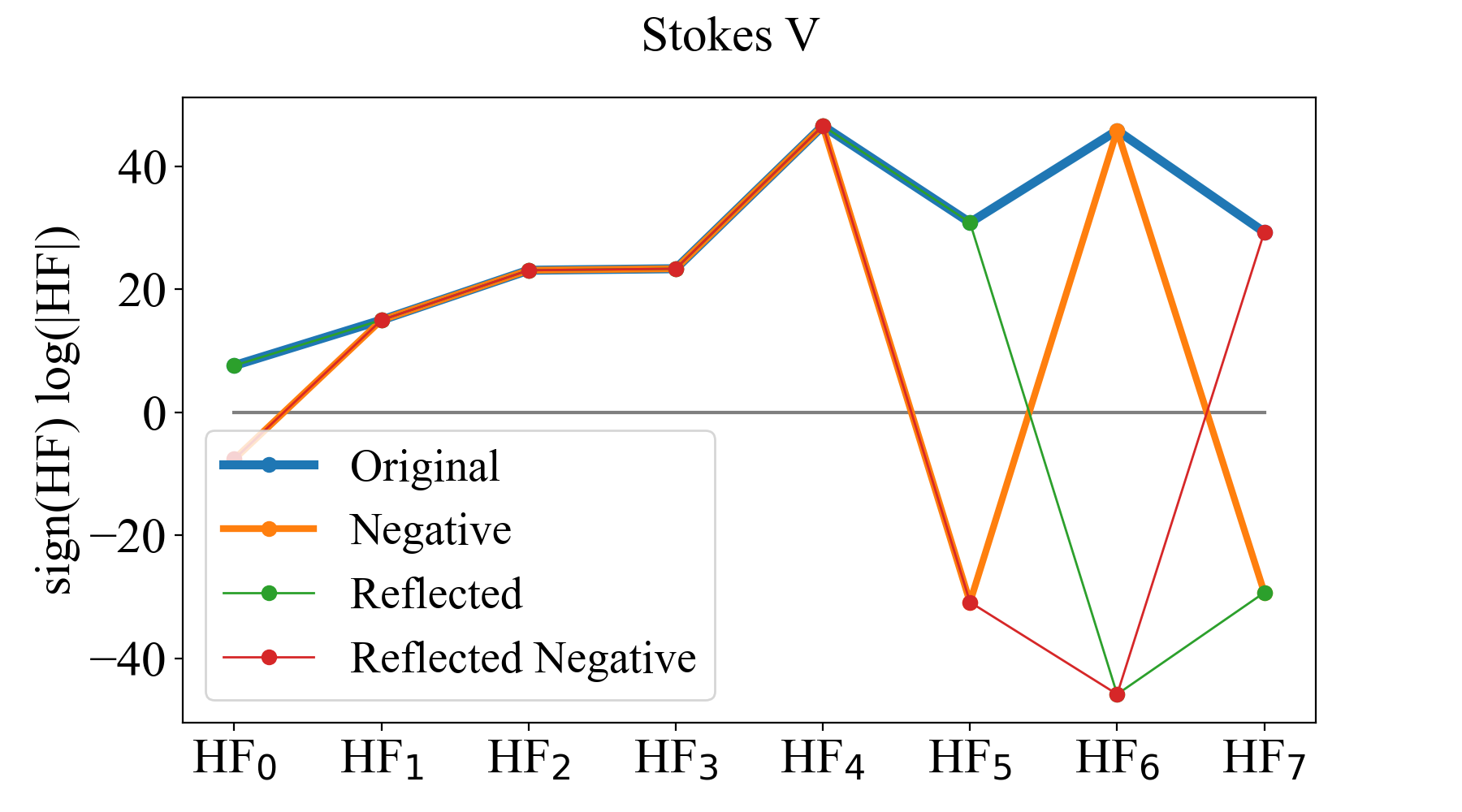}
    \end{subfigure}
    
\caption{Stokes $I$, $LP$ and Stokes $V$ images and invariants of a randomly selected frame from our library. Each column shows the image being modified by a different transformation. These are, in order: original, negative image, reflection, combination of negative image and reflection.}
\label{fig:inv_inversion}
\end{figure*}

We show how the HF invariants (Eq.~\ref{eq:hu_invariants}) behave under reflection and taking the negative of an image (Fig.~\ref{fig:inv_inversion}). Without loss of generality, we choose a random frame from our library.

When taking the negative of an image, there are three flips in the sign of the invariants. These are $HF_0,~HF_5,~HF_7$. The change in sign of $HF_0$ could be analogous to an inversion of the direction of rotation of the moment of inertia of a body about an axis. While a physical intuition for a flip in $HF_5$ is unclear, the change in $HF_7$ should indicate some degree of reflection with respect to the original image.

Reflection of an image produces two expected changes in the invariants: only $HF_6,~HF_7$ change sign. Since these exhibit antisymmetric properties under reflection, they are indicative of mirroring in the images. This could be because the centroid of the image shifted to the reflected position of that of the original image.

A combination of negative image and reflection causes three sign flips in $HF_0,~HF_5,~HF_6$. It is interesting that a sign change in $HF_7$ is cancelled out by both transformations.

\section{Minimized and maximized differences between invariants. }
\label{appdx:min_max_diff}

\begin{figure*}
\centering
\includegraphics[trim={0.0cm 0.0cm 0.0cm 0,0cm},clip=True,width=1.0\textwidth]{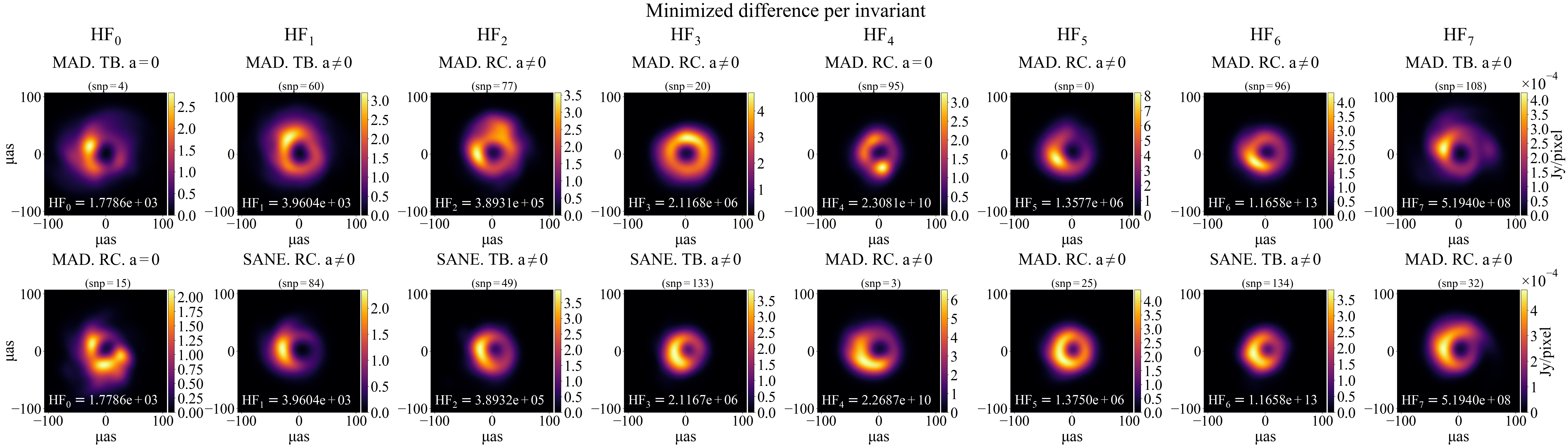} \\ \vspace{0.3cm} 
\includegraphics[trim={0cm 0cm 0cm 0cm},clip=True,width=1.0\textwidth]{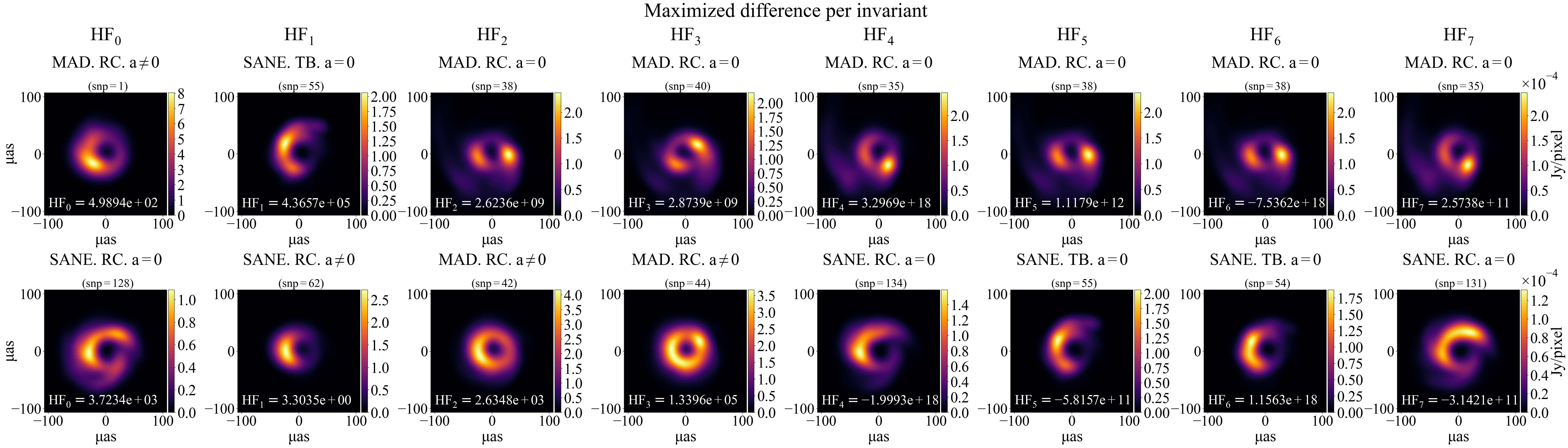}
\caption{Frames where the difference between the value of each HF invariant for Stokes I has been minimized (top) or maximized (bottom) considering the entire library. The difference between each IMI has been determined independently of the other seven. It is unclear how each HF invariant particularly grades the similarities and differences in the images.}
\label{fig:min_max_diff}
\end{figure*}

We explore the behaviour of HF invariants applied to our image library in an attempt to identify what particular image property is measured by each individual quantity (Eqs.~\ref{eq:hu_invariants}). 
Figure~\ref{fig:min_max_diff} shows the frames where the percentile differences (Eq.~\ref{eq:percentile_diff}) between each invariant have been minimized (top) or maximized (bottom) considering the all the frames in the entire library. Only Stokes $I$ has been considered, for illustrative purposes. Each difference for each IMI has been determined independently of the other seven. 
Unfortunately, no clear tendency is observed and the complicated functional form of the invariants hinders a clear interpretation of what each one is measuring. This is reflected by the similarities and differences seen in the images in all cases. 

\section{Details of the moment invariant distributions}
\label{appdx:distr_physics}

\begin{figure*}
\centering
    \begin{subfigure}{1.0\textwidth}
    \centering
    {\large Stokes~$I$} \\ 
    \includegraphics[trim={0cm 0cm 0cm 5.5cm},clip=True,width=1.0\columnwidth]{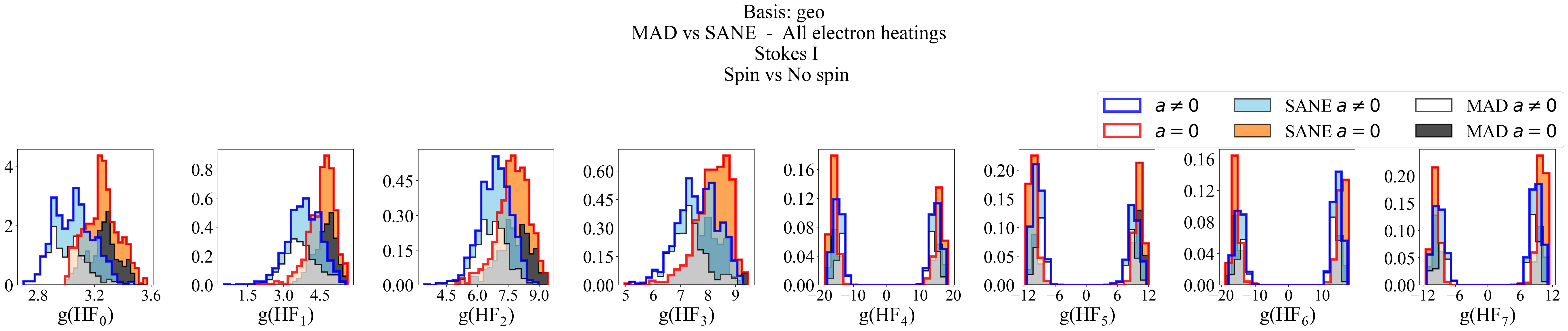}\\ \vspace{0.1cm}
    \includegraphics[trim={0cm 0cm 0cm 5.5cm},clip=True,width=1.0\columnwidth]{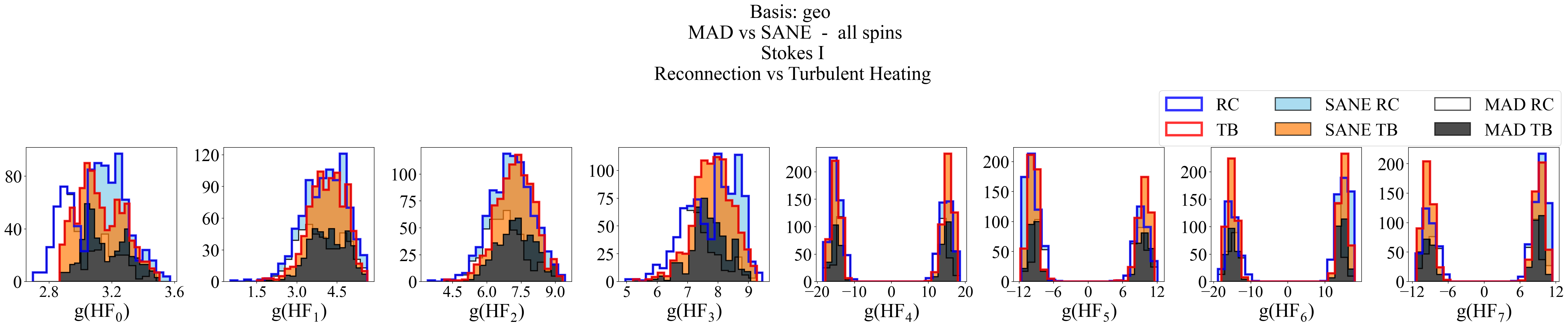}
    \end{subfigure}
\caption{Normalized stacked distributions of $g(HF_k)=\textrm{sign}(HF_k)\log|HF_k|$ of Stokes $I$ images as a function of various physical effects. Columns indicate a different HF invariant. Rows indicate: spin (top), electron heating (RC vs TB, bottom). }
\label{fig:distr_huI_sign}
\end{figure*}

\begin{figure*}
\centering
    \begin{subfigure}{1.0\textwidth}
    \centering
    {\large Linear Polarization} \\ 
    \includegraphics[trim={0cm 0cm 0cm 5.5cm},clip=True,width=1.0\columnwidth]{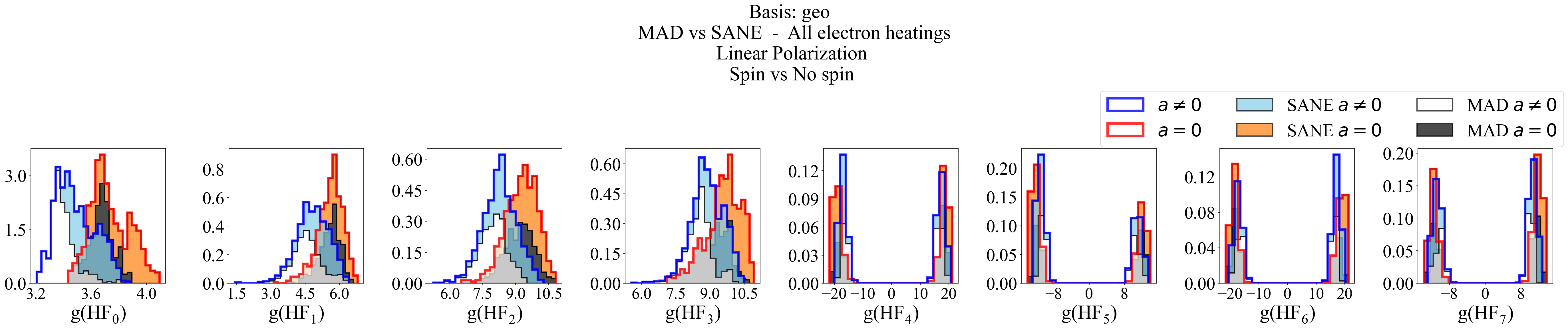}\\ \vspace{0.1cm}
    \includegraphics[trim={0cm 0cm 0cm 5.5cm},clip=True,width=1.0\columnwidth]{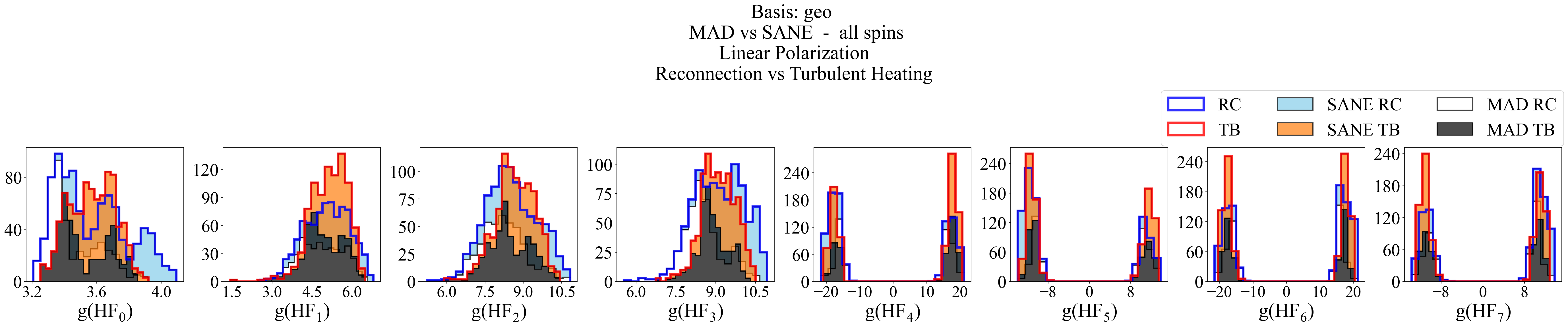}
    \end{subfigure}
\caption{Similar to Fig.~\ref{fig:distr_huI_sign} but for $LP$.}
\label{fig:distr_huLP_sign}
\end{figure*}

\begin{figure*}
\centering
\begin{subfigure}{1.0\textwidth}
    \centering
    {\large Stokes~$V$} \\ 
    \includegraphics[trim={0cm 0cm 0cm 5.5cm},clip=True,width=1.0\columnwidth]{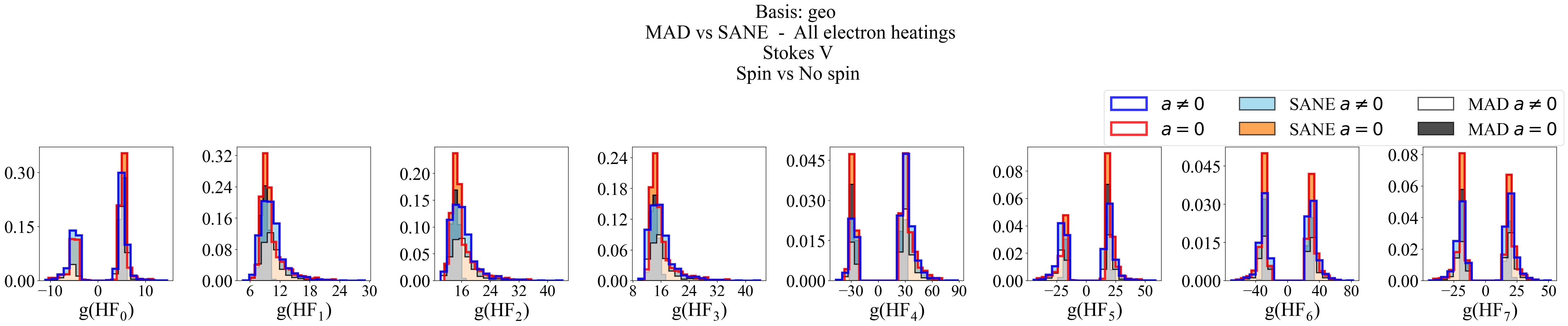} \\ \vspace{0.1cm}
    \includegraphics[trim={0cm 0cm 0cm 5.5cm},clip=True,width=1.0\columnwidth]{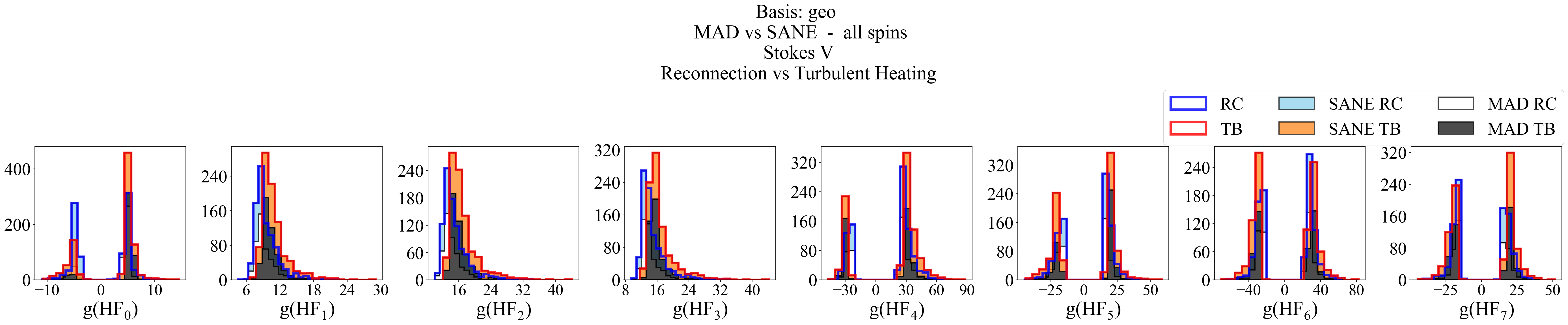}
    \end{subfigure}
\caption{Similar to Fig.~\ref{fig:distr_huI_sign} but for Stokes~$V$.}
\label{fig:distr_huV_sign}
\end{figure*}

Figs.~\ref{fig:distr_huI_sign},~\ref{fig:distr_huLP_sign},~\ref{fig:distr_huV_sign} 
show decomposition of the distributions of HF invariants as a function of different model parameters. This helps understand the origin of different peaks in the distributions.

\section{Application to jet-like images}
\label{appdx:jet_images}

\begin{figure*}
\centering
\includegraphics[trim={0cm 0cm 0cm 1.3cm},clip=True,scale=0.6]{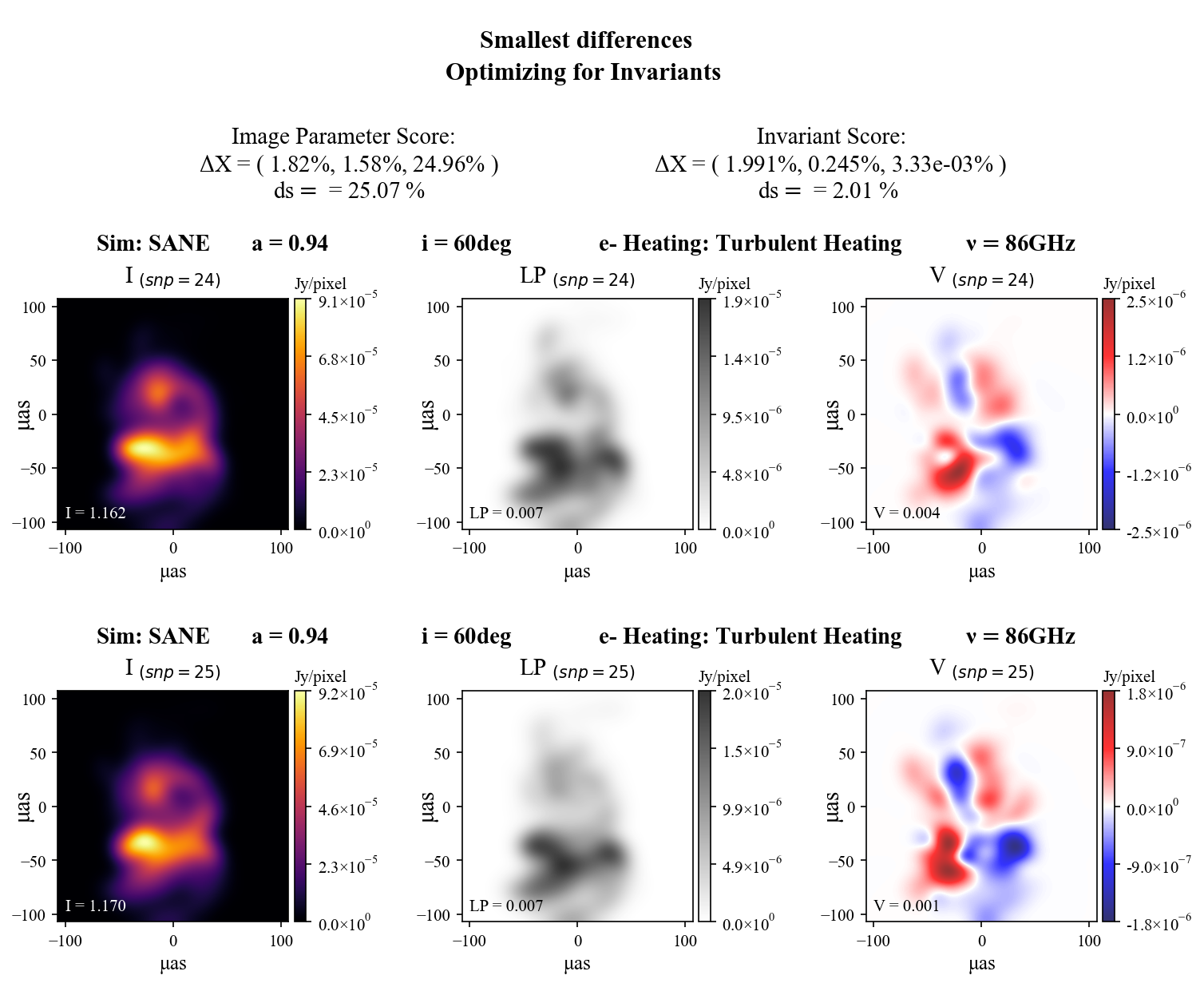}
    \caption{Closest frames in jet-like images of a SANE model optimized for the smallest difference between the HF invariants (Eq.~\ref{eq:deltaX_inv}). }
    \label{fig:closest_huinvs_jet}
\end{figure*}

We have applied our algorithm to identify the closest images from a set of non-ring like images in our library. We limited our choice to frames from the SANE TB high spin model at 86 GHz, which feature a very prominent jet.

In Fig.~\ref{fig:closest_huinvs_jet} we show the closest frames according to criterion 2 in (Eq.~\ref{eq:deltaX_inv}), where the frames between which the smallest joint difference from all the HF invariants is achieved. Because all the images from this sub-sample come from the same model, we have excluded all the possible pairs formed from comparing a frame with itself (150). In total, the algorithm selected the best possible combination of frames out of $150\times 150-150 = 22,350$ combinations. It can be seen that since the algorithm is not given the option to chose the same frame, it naturally chooses consecutive frames. This is in agreement with the high correlation between the snapshots of a model (see Section~\ref{sec:time}).

Even so, the invariants are sensitive enough to pick up on the fact it is not the same image and therefore, the $ds$ from invariants is still larger than the score obtained using image parameters.

\section{Application of scoring method to extended sample of images}
\label{appdx:extended_library}

\begin{table*}
	\centering
	\caption{Distance score applied to extended library.} 
	\label{table:extended_library}
	\begin{tabular}{lcccccccc} 
		\hline
        & $HF_0$ & $HF_1$ & $HF_2$ & $HF_3$ & $HF_4$ & $HF_5$ & $HF_6$ & $HF_7$\\
		\hline
 SANE & 1.726e3 & 1.632e4 & 1.750e7 & 2.648e8 & 1.590e16 & -3.827e9 & -8.489e15 & 1.681e10 \\
 MAD & 3.115e3 & 1.596e5 & 2.560e9 & 2.594e9 & -7.526e17 & 1.023e12 & -6.642e18 & 8.232e10 \\
 Donut & 2.5888e-4 & 7.265e-10 & 3.900e-13 & 4.170e-14 & -3.124e-27 & -1.067e-18 & -4.303e-27 & -1.767e-19 \\
 Eye & 6.061e-5 & 2.066e-11 & 3.908e-17 & 1.056e-15 & 2.022e-31 & 4.800e-21 & 7.208e-32 & 6.757e-23 \\
 Dog & 3.866e-4 & 7.101e-11 & 1.299e-12 & 3.093e-12 & 6.047e-24  & 2.488e-17 & -1.371e-24 & -3.882e-18 \\
  \hline
	\end{tabular}
\end{table*}

\begin{figure*}
\centering
\includegraphics[trim={0.0cm 0.0cm 0.0cm 0cm},clip=True,width=1.0\textwidth]{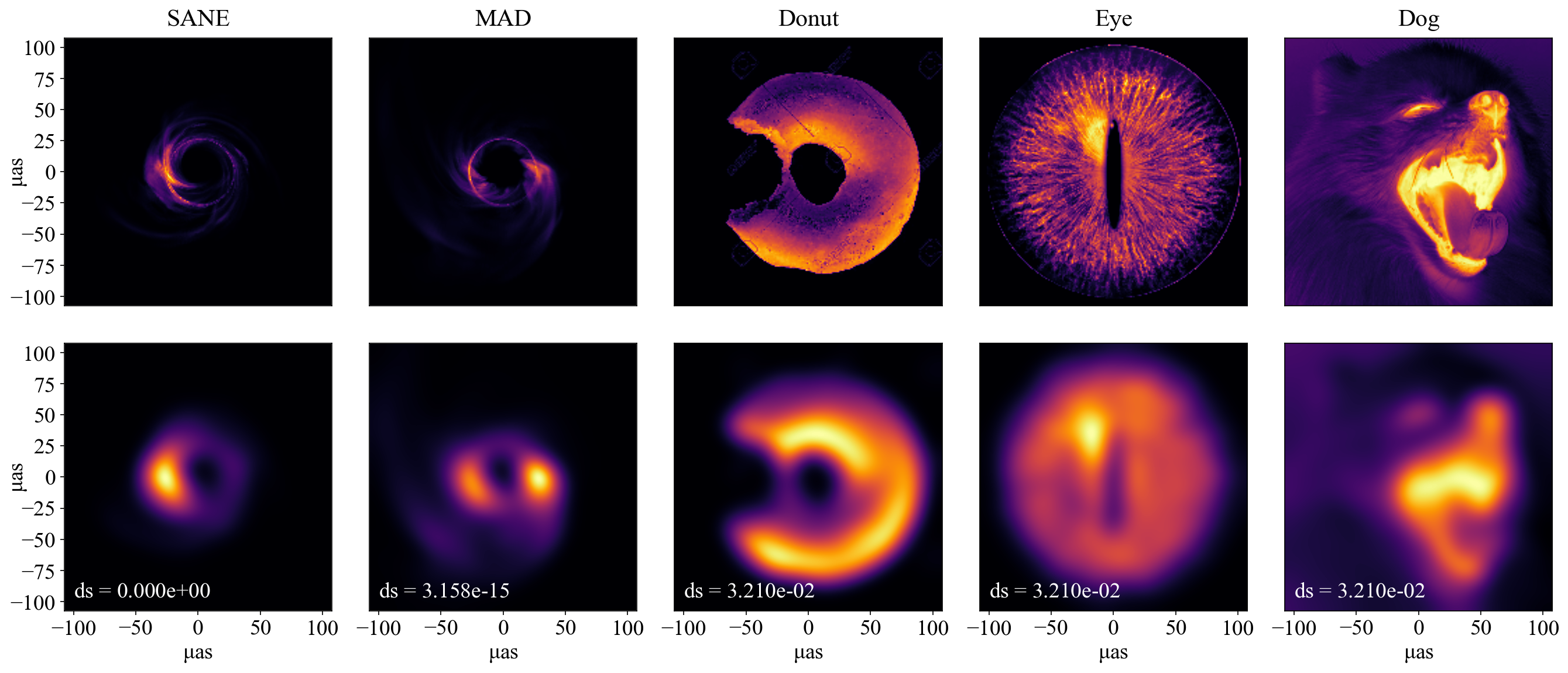}
\vspace{-0.3cm}
 \caption{Application of distance scoring algorithm to extended library. Top row: unblurred images. Bottom: blurred images by a $20~\mu$as gaussian. Taking as reference the SANE model, the images have been sorted according to their distance to it, with increasing values from left to right. $ds$ is indicated in the panels. }
    \label{fig:extended_library}
\end{figure*}

We have applied our scoring algorithm to a wider sample of images composed of two from our library and a few non-black hole images: a frame from a SANE $a=0.5$ TB model and one from a MAD $a=0$ RC (these have been chosen without loss of generality), a donut with a bite, an eye and a yawning dog (Fig.~\ref{fig:extended_library}, top row). The non-black hole images are in the public domain. 

All images have the same number of pixels $192\times 192$, and have been  blurred by a $20~\mu$as gaussian (the images have been assumed to cover the same field of view). Since the invariants are scale invariant, the is no need for normalization of fluxes.

Taking the SANE image as a reference, the bottom row of Figure~\ref{fig:extended_library} shows the images sorted according to their distance to it, with increasing values of $ds$ from left to right. The value of the distance score is indicated at the bottom left of each panel. The algorithm successfully finds the closest black holes among the collection of images. The $ds$ of the non-black hole images to the SANE appears to be the same for all. However, this is due to the invariants difference being divided by the difference between the maximum and minimum of the population of images (Eq.~\ref{eq:percentile_diff}). There is a seven order of magnitude difference between these values, which can be seen in Table~\ref{table:extended_library}.

\bsp	
\label{lastpage}
\end{document}